\pdfoutput=1
\documentclass[usenatbib]{mn2e}
\usepackage{apjfonts}
\usepackage{amssymb}
\usepackage{amsmath}
\usepackage{ctable}
\usepackage{url}
\usepackage{fixltx2e} 

\newcommand{\be}{\begin{equation}}
\newcommand{\ee}{\end{equation}}

\newcommand{\s}{{\,\rm s}}
\newcommand{\cm}{{\,\rm cm}}
\newcommand{\etal}{et al.}

\newcommand{\msun}{M_{\sun}}

\newcommand{\msunyr}{M_{\sun}\,{\rm yr^{-1}}}
\newcommand{\tauavg}{\tau_{0}}
\newcommand{\movieurl}{\url{https://www.cfa.harvard.edu/~phopkins/Site/Research.html}}

\newcommand{\scaleup}{}

\newcommand\plotone[1]
 {\centering \leavevmode \includegraphics[width={0.99\columnwidth}]{#1}}

\newcommand{\plotside}[1]
 {\centering \leavevmode \includegraphics[width={0.95\textwidth}]{#1}}
\newcommand{\plotsidesmall}[1]
 {\centering \leavevmode \includegraphics[width={0.90\textwidth}]{#1}}

\newcommand{\plotsidesmallest}[1]
 {\centering \leavevmode \includegraphics[width={0.7\textwidth}]{#1}}
\newcommand{\plotsidesmallestest}[1]
 {\centering \leavevmode \includegraphics[width={0.65\textwidth}]{#1}}
\newcommand{\acknowledgments}{\begin{small}\section*{Acknowledgments}\end{small}}
\newcommand\altaffilmark[1]{$^{#1}$}
\newcommand\altaffiltext[1]{$^{#1}$}
\title[Self-Regulated Star Formation in Galaxies]{Self-Regulated Star Formation in Galaxies via Momentum Input from Massive Stars}

\author[Hopkins \etal]{
\parbox[t]{\textwidth}{ 
Philip F.~Hopkins\thanks{E-mail:phopkins@astro.berkeley.edu}\altaffilmark{1},
Eliot Quataert\altaffilmark{1}, \&
Norman Murray\altaffilmark{2,3} 
}
\vspace*{6pt} \\
\altaffiltext{1}{Department of Astronomy and Theoretical Astrophysics Center, University of California Berkeley, Berkeley, CA 94720} \\
\altaffiltext{2}{Canadian Institute for Theoretical Astrophysics, 
60 St.\ George Street, University of Toronto, ON M5S 3H8, Canada} \\
\altaffiltext{3}{Canada Research Chair in Astrophysics} 
}

\date{Submitted to MNRAS, January, 2011}
\begin{document}
\maketitle
\label{firstpage}

\begin{abstract}

Feedback from massive stars is believed to play a critical role in shaping the galaxy mass function, the structure of the interstellar medium (ISM), and the low efficiency of star formation, but the exact form of the feedback is uncertain. In this paper, the first in a series, we present and test a novel numerical implementation of stellar feedback resulting from momentum imparted to the ISM by radiation, supernovae, and stellar winds. We employ a realistic cooling function, and find that a large fraction of the gas cools to $\lesssim 100$ K, so that the ISM becomes highly inhomogeneous. Despite this, our simulated galaxies reach an approximate steady state, in which gas gravitationally collapses to form giant `molecular' clouds (GMCs), dense clumps, and stars; subsequently, stellar feedback disperses the GMCs, repopulating the diffuse ISM. This collapse and dispersal cycle is seen in models of SMC-like dwarfs,  the Milky-Way, and  $z \sim 2$ clumpy disk analogues. The simulated global star formation efficiencies are consistent with the observed Kennicutt-Schmidt relation. Moreover, the star formation rates are nearly {independent} of the numerically imposed high-density star formation efficiency, density threshold, and density scaling. This is a consequence of the fact that, in our simulations, star formation is regulated by stellar feedback limiting the amount of very dense gas available for forming stars. In contrast, in simulations without stellar feedback, i.e., under the action of only gravity and gravitationally-induced turbulence, the ISM experiences runaway collapse to very high densities. In these simulations without feedback, the global star formation rates exceed observed galactic star formation rates by $1-2$ orders of magnitude, demonstrating that stellar feedback is crucial to the regulation of star formation in galaxies.

\end{abstract}

\begin{keywords}
galaxies: formation --- galaxies: evolution --- star formation: general --- cosmology: theory
\end{keywords}

\section{Introduction}
\label{sec:intro}

Feedback from massive stars plays a critical role in the evolution of galaxies. Cosmological models of galaxy evolution generically find that, without strong stellar feedback, the net stellar mass formed from cooled baryons exceeds that observed by an order of magnitude or more, particularly in lower mass halos \citep[e.g.][and references therein]{katz:treesph, somerville99:sam,cole:durham.sam.initial, springel:lcdm.sfh,keres:fb.constraints.from.cosmo.sims}.  Related problems exist on smaller scales within galaxies.   The observed relationship between star formation rate density and gas surface density -- the Kennicutt-Schmidt (KS) law -- implies that star formation is slow   averaged over galaxies as a whole:  the gas consumption timescale is $\sim50$ dynamical times \citep{kennicutt98}, much longer than the naive estimate of $\sim$ a few dynamical times one might expect in self-gravitating gas.  Similar gas consumption times are found even in dense regions in galaxies (e.g., \citealt{krumholz:sf.eff.in.clouds}; 
but see also \citealt{murray:2010.sfe.mw.gmc,schruba:2010.mol.gas.depletion.time,feldmann:2011.gmc.sfe.eff.vs.time}).    
Moreover, observations in the Milky Way and nearby galaxies show that individual giant molecular clouds (GMCs) -- the sites of star formation -- convert only a few percent of their mass into stars during their lifetimes \citep{zuckerman:1974.gmc.constraints,williams:1997.gmc.prop,evans:1999.sf.gmc.review,
evans:2009.sf.efficiencies.lifetimes}.   One of the leading explanations for this low star formation efficiency is that stellar feedback disrupts GMCs once enough stars have formed.  

Numerical simulations of isolated galaxies and galaxy mergers, as well as cosmological ``zoom-in'' simulations of individual halos, can now reach the resolution required to resolve the formation of GMCs, $\sim 1-100$ pc 
\citep[see e.g.][]{bournaud:2010.grav.turbulence.lmc,saitoh:2008.highres.disks.high.sf.thold,
dobbs:2011.why.gmcs.unbound,tasker:2009.gmc.form.evol.gravalone}
(note that GMCs in massive gas-rich galaxies are $\sim$ kpc in size, significantly larger than in the Milky Way). 
If simulations do not, however, include physics that disrupts GMCs, they do not have a physically self-consistent model of the interstellar medium (ISM) on such scales. All of the gas will be unrealistically locked up in dense gaseous/stellar clusters, instead of being recycled back into the more diffuse ISM.  
Given resolution limitations, most recipes in 
galaxy and cosmological-scale simulations have been developed to treat star formation and feedback 
in a restricted  ``sub-grid'' manner. 
However, without more detailed models of this physics, it is difficult to assess how appropriate the sub-grid prescriptions are for different galaxy types. Moreover, the assumptions of such models 
break down and are no longer meaningful at the spatial ($\lesssim\,$pc) or time evolution ($\lesssim\,$Myr) scales of individual GMCs and ISM sub-structure. 
In particular, whenever a numerical simulation has the resolution to resolve the {\em formation} of bound gaseous structures like GMCs, we believe that it is equally critical to include physics that can potentially {\em disrupt} such GMCs.     

Protostellar jets, HII regions, stellar winds, radiation pressure from young stars, and supernovae all appear to be important sources of feedback and turbulence in the ISM of galaxies.  In regions of low mass star formation it is likely that protostellar jets dominate, but for the ISM as a whole massive stars are the most important sources of feedback.   In relatively low density gas, heating by photoionization, stellar winds, and supernovae is critical \citep{mckee.ostriker:ism,matzner02}.   For denser gas, however, which often corresponds to most of the mass in a galaxy, energy deposition is ineffective; the cooling time ($\tau_{cool}= kT/\Lambda n\approx 3000(T/10^4K)(1\cm^{-3}/n)$ yrs, where $\Lambda\approx10^{-23}\,{\rm erg}\cm^3\s^{-1}$ is the cooling function) is short compared to the dynamical time for all but the lowest density gas, so the energy deposited into the gas by stellar feedback is rapidly radiated away.  Even in the Milky Way, the hot ISM contributes only $\sim 10 \%$ to the total ISM pressure \citep{boulares90}. In contrast, the momentum supplied by stellar luminosity, winds, and supernovae cannot be radiated away, and is the most important source of feedback for dense gas in galaxies (e.g., \citealt{murray:molcloud.disrupt.by.rad.pressure}).  

Although it is widely believed that stellar feedback is critical for understanding the self-regulation of star formation within galaxies, and for the cosmological evolution of galaxies themselves, it is quite challenging to treat this 
in galaxy-scale simulations, especially with the computational limitations faced by previous generations of simulations. 
As a result, many simulations have made the problem tractable by adopting effective equation of state models 
 in which feedback processes are treated implicitly \citep[e.g.][]{springel:multiphase,
 teyssier:2010.clumpy.sb.in.mergers}, 
accounting for the (un-resolved) multi-phase turbulent structure of the ISM with an ``effective'' high sound speed. 
Unfortunately, in this case many of the net effects of stellar feedback are then put in by hand -- one cannot predict, e.g., either how efficient feedback is in different systems or whether stellar feedback drives galactic winds. More broadly,  without simulations that explicitly model feedback, it is difficult to evaluate the accuracy of the various subgrid treatments in the literature.    

Galactic-scale simulations that do include stellar feedback explicitly have often been forced to alter the physics in significant ways in order to obtain a desired result.    The most common treatment is to only include thermal gas heating from supernovae.  However, thermal feedback is very inefficient in the dense regions where star formation occurs, and in the ISM more broadly in starbursts and gas-rich high-redshift galaxies. 
These problems are compounded when simulations cannot resolve the ISM phase structure, and smooth together dense GMCs and diffuse gas into a single average density (greatly increasing/decreasing the cooling time in dense/diffuse gas, respectively). 
For this reason, in order to make supernova feedback have any significant effect (even in MW-like galaxies), simulators often ``turn off'' cooling (often along with star formation and/or 
other hydrodynamic processes) for an extended period of time, much longer than $\tau_{cool}$ \citep[cooling is typically suppressed for $\sim10^{7}-10^{8}\,$yr, i.e., for a time comparable to a galaxy dynamical time and $\sim10^{4}$ times longer than the actual cooling time at the same density; see e.g.][]{thackercouchman00,governato:disk.formation,brook:2010.low.ang.mom.outflows}. 
Other models explicitly disable certain interactions between gas flagged as ``cold'' and ``hot'' or deposit feedback energy 
in a non-cooling reservoir that serves to move gas from cold to hot phases \citep{scannapieco:fb.disk.sims}. 
Even with these adjustments, many such models have found it difficult to drive winds and suppress star formation 
at the level needed to explain the galaxy mass function (especially 
at low masses) and observed star formation efficiencies
\citep[see e.g.][and references therein]{guo:2010.hod.constraints,powell:2010.sne.fb.weak.winds,
brook:2010.low.ang.mom.outflows,nagamine:2010.dwarf.gal.cosmo.review}.

Simulations with supernova feedback that do not ``turn off'' cooling
have found that galactic outflows can only be driven if additional physics is included. For example, \citet{ceverino:cosmo.sne.fb} were able to drive galactic winds by requiring that supernovae explode well outside of the GMCs in which they formed. However (as they acknowledge), although this may well be important for galactic winds, it leaves the problem of locally preventing runaway collapse of dense star forming regions.   

The inefficiency of supernova heating in dense gas is physically correct.\footnote{It is, of course, true that simulations do not resolve the full multi-phase ISM into which supernovae propagate and that this can enable supernova energy to propagate to larger distances.}    It is thus by no means clear that turning off cooling is an appropriate resolution of the `problem' that supernova remnants cool!   Instead, we believe that this points to the importance of including the momentum supplied by stellar feedback processes.   This momentum input into the ISM can drive strong turbulence and can itself contribute to unbinding gas from galaxies, even in the limit of very rapid cooling \citep{murray:momentum.winds}.    To date, however, this has only been treated in a phenomenological way, given the limited resolution of previous simulations.   In particular, in a widely used implementation in the Gadget SPH code, gas particles are "kicked" into a ``wind'' at a rate proportional to either the star formation rate or young stellar mass, with the wind velocity set by hand as a constant or a multiple of the galaxy escape velocity \citep{springel:multiphase,oppenheimer:metal.enrichment.momentum.winds,
sales:2010.cosmo.disks.w.fb,genel10,dalla-vecchia:2008.iso.gal.wwo.freestream.winds}.   All hydrodynamic interactions (e.g., shocks and pressure forces) for the ``wind particles'' are turned off until they escape the galaxy \citep[and when this is not done, the effects of winds are substantially suppressed;][]{sales:2010.cosmo.disks.w.fb}. This model is useful for studying the impact of galactic outflows on the intergalactic medium and the galaxy mass function but it is clearly limited (especially within galaxies) and cannot predict the nature and origin of these winds.

Motivated by these considerations, this is the first in a series of papers studying stellar feedback in numerical models of galaxies and the resulting implications for problems such as the origin of galactic winds, the physics of gas inflow in galaxy mergers, and the properties of the ISM in high redshift galaxies.\footnote{\label{foot:url}Movies of these
  simulations are available at \movieurl}     Ultimately, we will present results that include simple models of supernova heating, HII regions, and radiation pressure from massive stars (produced by the absorption and scattering of UV and IR radiation on dust).    Feedback from a central active galactic nucleus may also be important for understanding some aspects of star formation in galaxies -- particularly the cessation of star formation in massive galaxies --  but this is a separate problem that we do not consider in this paper.

It is still not currently computationally feasible to include all of the physics of stellar feedback in simulations that focus on galactic scales.   The methods we develop therefore still rely on sub-grid models, but at the sub-cluster or sub-GMC scale, as opposed to the galaxy scale.   The fact that different feedback processes dominate under different physical conditions (e.g., density) highlights the importance of including a range of physical processes when studying the effects of stellar feedback on galaxies and galaxy formation.   Nonetheless, in this paper, we focus on isolated (non-cosmological) galaxies and only include feedback by momentum deposition from massive stars.   Our motivation for doing so is several-fold.   First, our model for momentum-deposition is sufficiently different from existing treatments of stellar feedback in the literature that it requires a detailed explanation.  More importantly, however, we show that this simple model is, by itself, able to explain the Kennicutt-Schmidt relation and the low star formation efficiency in galaxies.   Moreover, the star formation rates in our model galaxies typically change by less than a factor of $\sim 2$ when we include additional feedback processes (though other properties of the galaxies can change substantially, such as the morphology and phase-structure of the ISM -- this is particularly true for low mass galaxy models).   

The remainder of this paper is organized as follows.   In \S \ref{sec:sims} we describe our method of implementing feedback due to the injection of momentum by young, massive stars.   The Appendix contains tests varying some of the parameters of our numerical method.   The galaxy models we study are described in Table \ref{tbl:sim.ics} and \S \ref{sec:ICs}.   We then summarize our key results on the star formation histories and structural properties of our model galaxies (\S \ref{sec:results:global}).   In \S \ref{sec:results:sfh} we show that these results do not depend strongly on the physics of star formation at high densities, the uncertain feedback parameters, and numerical resolution.   We then show that our model galaxies are  consistent with the observed Kennicutt-Schmidt relation (\S \ref{sec:results:ks}).   Finally, in \S \ref{sec:discussion} we summarize our results and discuss their implications.  

\vspace{-0.3cm}
\section{Methodology}
\label{sec:sims}

The methods we present are general and can be implemented in both Eulerian and Lagrangian simulations.   The specific simulations we carried out were performed with the parallel TreeSPH code {\small GADGET-3} \citep{springel:gadget}, based on a conservative formulation of smoothed particle hydrodynamics (SPH), which conserves
energy and entropy simultaneously even when smoothing lengths evolve
adaptively \citep[see e.g.,][]{springel:entropy,hernquist:sph.cautions,oshea:sph.tests}.
The detailed numerical methodology is described in
\citet{springel:gadget}, \citet{springel:multiphase}, and
\citet{springel:models}.  Our simulations include stars, dark matter, and gas, 
with new implementations of stellar feedback; we describe the salient features of this additional physics below.   These calculations do not include models of black hole growth and feedback. 

\subsection{Cooling and Star Formation}
\label{sec:stellar.model}



In order to resolve the formation of very dense clumps, we extend the 
standard atomic+metal line cooling 
curves in GADGET (which cut off when the gas becomes 
neutral at $<10^{4}\,$K) to allow cooling by fine structure lines. 
Specifically we tabulate the cooling function $\Lambda(T)$ from $1-10^{4}\,$K with CLOUDY, for a medium with 
density $n=1\,{\rm cm^{-3}}$, solar abundances, and with an 
ionizing background matching that at $z=0$.\footnote{Recalibrating 
our ``baseline'' $\Lambda(T)$ at $n=100\,{\rm cm^{-3}}$ gives indistinguishable results. 
The difference (modulo the standard $n^{2}$ dependence) is much smaller than 
more dramatic cooling curve variations we consider among other numerical 
tests in \S~\ref{sec:appendix}, which all produce nearly identical results because, in all 
cases, the cooling time is much less than the dynamical time.} 
This is similar to the approach in a number of other simulations 
\citep[see e.g.][]{robertson:2008.molecular.sflaw,wise:2008.first.star.fb,ceverino:cosmo.sne.fb}
and gives identical results to the tabulated $\Lambda(T)$ 
presented in \citet{sanchez-salcedo:2002.coolingcurves}. 
We are not attempting to follow the ISM chemistry and thus ignore the dependence of the cooling on abundance and radiation field.
For our problems of interest, the cooling rates even at these low temperatures are uniformly much shorter than the dynamical times in all the systems we model; therefore, even large (factor $\sim5$) changes in the cooling curve make no significant difference to our conclusions (we have checked this explicitly).

Because we allow cooling to very low temperatures, we also must account for finite simulation resolution by including a pressure floor 
to prevent artificial numerical fragmentation when the Jeans 
mass is not resolved \citep{truelove:1997.jeans.condition}. We adopt the prescription in 
\citet{robertson:2008.molecular.sflaw}, which ensures that the Jeans length is 
always resolved with $N_{\rm Jeans}$ smoothing lengths. This density-dependent pressure floor is
\begin{equation}
\label{eqn:Pjeans}
P_{\rm Jeans} \approx 1.2\,N_{\rm Jeans}^{2/3}\,\gamma^{-1}\,G\,h_{\rm sml}^{2}\,\rho^{2}
\end{equation}
where $\gamma=5/3$, $\rho$ is the local density, and $h_{\rm sml}$ the smoothing length. 
We typically adopt $N_{\rm Jeans}=10$, but have experimented with $N_{\rm Jeans} = 4-15$ and find similar results 
(the morphologies, SFRs, and Schmidt-Kennicutt relations are indistinguishable; see Appendix~\ref{sec:appendix}). 
We make one small modification to the prescription in \citet{robertson:2008.molecular.sflaw}, which is to 
track the numerical pressure floor separately so that it enters into the momentum equations, but 
does not explicitly change the gas temperature (relevant, e.g., for determining the cooling function). 
This is a standard approach in high-resolution simulations 
\citep[see e.g.][]{teyssier:2010.clumpy.sb.in.mergers}.   At the typical resolution we adopt, 
the pressure provided by equation~\ref{eqn:Pjeans} is much less than the turbulent pressure 
resulting from our feedback model (by a factor of $\sim10^{2}-10^{4}$); 
only when we turn off feedback entirely is the ISM pressure resolution-limited. 

In our simulations, stars are assumed to form from dense gas with a constant efficiency $\epsilon$ per free-fall time $t_{\rm ff}=\sqrt{3\pi/32\,G\,\rho}$, above some minimum threshold $\rho_{0}$, i.e.\, 
\be
\dot{\rho}_{\ast} = \frac{\epsilon\,\rho}{t_{\rm ff}} \propto \rho^{3/2} \ \ {\rm for} \ \ \rho > \rho_0 \label{eqn:sfr}\ee  
This is numerically implemented by turning gas particles into stars stochastically
following the calculated SFR (probability $p=1-\exp{(-\dot{\rho}_{\ast}\,dt/\rho)}$, 
where $dt\lesssim100-1000\,$yr 
is the simulation timestep and also represents how frequently these values are updated).
Because we wish to resolve the dense regions of 
star formation, we typically set $n_{0}=100\,{\rm cm^{-3}}$, 
characteristic of large GMCs. 
The efficiency $\epsilon$ is empirically measured in dense star-forming 
regions to be $\approx1-2\%$, roughly constant over a wide 
range of densities $\sim10-10^{6}\,{\rm cm^{-3}}$ \citep{krumholz:sf.eff.in.clouds};  we adopt a canonical value of $\epsilon=1.5\%$ \citep[see also][]{leroy:2008.sfe.vs.gal.prop}. 
We discuss variations about these fiducial choices in \S \ref{sec:results:sfh:sflaw} below. 

\subsection{Stellar Feedback}
\label{sec:stellar.model:fb}

For the reasons summarized in \S \ref{sec:intro}, we model stellar feedback by depositing momentum into the gas around young star clusters.   This in turn drives strong turbulence in the ISM.  For standard IMFs (e.g., \citealt{kroupa:imf}), the momentum supplied to the ISM by stellar winds, supernovae, and the luminosity of young stars are all comparable \citep{starburst99,murray:momentum.winds}.    If supernovae undergo a significant Sedov-Taylor phase, the P-dV work done can increase their momentum by a factor of $\sim 10$ (e.g., \citealt{thornton98}).   Likewise, if the ISM is optically thick to the infrared radiation produced when dust reradiates stellar UV photons, the radiation energy density builds up, increasing the radiation pressure force by a factor of the infrared optical depth $\tau_{IR}$.   Modeling these processes in detail is a daunting task and one that is beyond the scope of this paper.   Instead, we explore the general properties of models in which turbulence driven by momentum-deposition is the dominant stellar feedback mechanism.   This is a plausible approximation even in the Milky Way, since the hot ISM contributes only $\sim 10 \%$ to the ISM pressure \citep{boulares90}.   Moreover,  in the well-studied star forming region 30 Doradus in the LMC, {observations} directly implicate radiation pressure as the dominant mechanism of stellar feedback \citep{lopez:2010.stellar.fb.30.dor}. 
We stress, however, that this is not intended to represent a complete model of stellar feedback and the ISM; in future work, we will study how galaxy properties are further modified with the addition of other mechanisms such as SNe and stellar wind shock-heating and mass loss, and photoionization heating.

In order to make our simplified feedback model as realistic as possible, we implement the feedback so that it is explicitly associated with young star clusters.  We do so by identifying star-forming clumps and depositing momentum into the surrounding gas radially away from the star clusters.   In the following subsections we describe the key steps in this method.

\vspace{-0.2cm}

\subsubsection{Star-Forming Clumps: Identification}
\label{sec:stellar.model:clumpids}

The first step is to identify star forming regions or nascent star clusters in the 
simulation. Starting from each gas particle, we identify the nearest dense gas 
``clump'' by iteratively performing a friends-of-friends search with an adaptive 
linking length. Specifically, we search over all 
gas particles within a radius $N_{\rm sml}\,h_{\rm sml}$  (with typical $N_{\rm sml}=3$) 
of the initial particle to find that particle with the highest local density, and iterate either until a higher-density 
neighbor is not found or until some maximum cutoff is reached.
For the latter we adopt a maximum of $20$ iterations or a distance $>20$ times 
the initial particle $h_{\rm sml}$ (in practice, this limit is rarely reached, 
but is necessary to prevent cases where, e.g., the linking chain might traverse
a large fraction of the length of a spiral arm). 
Some care is needed in choosing the appropriate value of $N_{\rm sml}$, 
based on the physical scales that are or are not resolved in a given 
simulation -- for our resolutions, $N_{\rm sml}<1$ will simply return 
the local gas particle, and $N_{\rm sml}\gg5$ tends to over-link clumps in 
dense regions such as spiral arms and galactic nuclei. 
Our experiments show that within the range $N_{\rm sml}\sim1.5-4$ the identification of the peak local density is converged for $>90\%$ of all ``clumps'' (with the remainder making little difference 
in global quantities, as we show explicitly in the Appendix); the density peaks identified in this way also agree well with visual identification of overdensities.  We thus adopt a canonical value of $N_{\rm sml}=3$.  

This friends-of-friends search defines the star-forming clump of which the 
initial gas particle is a member.\footnote{We have also
experimented with using the center of stellar light or the stellar density peak as the location of the clump center (see the Appendix for details). In the systems we model here, there is no 
detectable difference between these choices and our fiducial choice of centering the clump on the peak gas density. However, the distinction between peak gas and stellar quantities could be more important in systems where the main sequence lifetime exceeds 
the dynamical time (e.g., galactic nuclei) and massive stars may wander away from their natal GMCs.}
   The distance between the original particle and the
clump center (the clump density peak) defines the ``clump radius'' $R_{\rm clump}$  (if this distance is less than twice the initial smoothing length, we set it to this minimum value, since the ``clump'' is 
effectively unresolved). The enclosed ``clump mass'' in gas ($M_{\rm clump,\, gas}$) 
or stars ($M_{\rm clump,\, \ast}$) are defined as the mass of each component
within a distance of $R_{\rm clump}$ of the clump center.

\subsubsection{Momentum-Loading}
\label{sec:stellar.model:momentum}

In our model, stellar feedback is tied to the properties of the stars in the stellar cluster associated with a given gas particle.   Moreover, we only apply the feedback to gas particles that are within 
$3\,h_{\rm sml}$ of a young star particle (typically $\lesssim10\,$pc). This helps ensure that the feedback is spatially correlated with young stars.\footnote{Because the momentum deposition falls off for gas further from the stars, formally extending this to all of the gas makes no difference to our results.} 
We now motivate our implementation in terms of feedback by radiation pressure on dust grains.

At each timestep, we identify the stars (of those formed since the beginning of the simulation) 
within the previously-identified clump,  and sum their bolometric luminosity, which is a function of the star's
(known) stellar age
\be
\label{eqn:Ltot}
L_{\rm tot}(<R_{\rm clump}) = \sum_{i}^{r<R_{\rm clump}}
{{\Bigl [}\frac{L}{M}({{\rm age}_{i}}){\Bigr ]}\,\times M_{\ast,\,i}}
\ee
We tabulate $L_{\ast}/M_{\ast}$ as a function of 
age using a Starburst99 \citep{starburst99} single stellar population 
with a \citet{kroupa:imf} IMF at 
solar metallicity \citep[this time dependence can be important on GMC timescales, 
in contrast to models where all energy is coupled instantaneously; see 
e.g.][]{slyz:2005.kpc.box.sf.feedback}. 
Given the uncertainties in the mass-loading factors below, 
and the fact that our initial conditions are all relatively evolved systems, 
it makes little difference whether we explicitly allow for a metallicity 
dependence.\footnote{We neglect for now the fact that at extremely high resolution, a $<100\,\msun$ 
``star particle'' may not completely sample the stellar IMF, and simply take the average 
$L_{\ast}/M_{\ast}$ for the particle age. 
Since we focus on galaxy-average quantities, this is probably not a large 
uncertainty. But in low-mass clusters and GMCs, a more accurate model -- for example 
one which the stellar mass range of 
each particle is sampled stochastically from the IMF, as discussed in \citet{mashchenko:2008.dwarf.sne.fb.cusps} -- 
could give interesting differences.} 
Assuming that the stellar flux is equally distributed among all of the
gas within $R_{\rm clump}$, we obtain the luminosity $L_j$ incident
on the particle in question, which has a mass $M_{{\rm gas},\,j}$:
\be 
\label{eqn:Lj}
L_{j} = L_{\rm tot}(<R_{\rm clump}) \frac{M_{{\rm gas},\,j}}{M_{\rm tot,\,gas}(<R_{\rm clump})}
\ee
Because the luminosity incident on a particle in this simple formulation depends on the light-to-mass ratio of the surrounding stars,  we find that our results are relatively insensitive 
to whether we use the starlight within $R_{\rm clump}$ or some multiple of this radius.

Given the incident luminosity, we take the rate of momentum deposition in the gas to be
\be
\label{eqn:mom.vs.L}
\dot{p}_{j} = (1 + \eta_{p}\tau_{\rm IR})\,\frac{L_{j}}{c}.
\ee
This is the core equation of our feedback model (with typical values of $\tau_{\rm IR}$ in Fig.~\ref{fig:turb.highz.demo}).
This force is directed radially away from the clump center (i.e.\ along the vector ${\bf R}_{\rm clump}$). If the particle $j$ itself is the clump center, the direction of the force is randomly chosen isotropically.

The first factor of $L_j/c$ in equation \ref{eqn:mom.vs.L} represents the momentum imparted as the optical-UV photons emitted by massive stars are absorbed by dust, which re-radiates in the IR.   The factor of $\tau_{\rm IR} L_j/c$ accounts for the momentum imparted by the total number of IR photons absorbed/scattered within the gas parcel.   Note that $\tau_{\rm IR}$ is the optical depth through the clump, not the optical depth of the given gas particle.   It is the former that sets the total momentum supplied to the gas.   Finally, equation \ref{eqn:mom.vs.L} includes a dimensionless factor $\eta_{p} \sim 1$ that accounts for other sources of momentum and uncertainties introduced by our simplified treatment.   Note, e.g., that we do not explicitly include the momentum deposited by stellar winds and supernovae separately from that due to the radiation of massive stars;  $\eta_p \gtrsim 1$ crudely accounts for these additional contributions.\footnote{We have considered experiments where we include a separate, explicit $\dot{p}_{j}$ term for the direct momentum flux 
from stellar winds and SNe ejecta, with both tabulated in STARBURST99 as a function of stellar age. The absolute magnitude of these is, for a constant SFR, $\sim L/c$. We find this make no difference compared to 
equivalent variation in $\eta_{p}$.}
 On the other hand, $\eta_{p}\lesssim1$ might be appropriate if photons efficiently leak out through holes in the gas distribution (see Appendix~\ref{sec:appendix:leakage}).

Why do we associate the feedback with the clump and direct it 
from that center of density, as opposed to simply identifying it with each star individually? 
Recall, we are modeling the effects of radiation pressure in the limit in which the gas is 
at least somewhat optically thick. 
If the UV/optical photons could free-stream, then the appropriate sources would indeed 
be each star particle. 
However, if a number of stars are embedded in a gas clump, 
then in the limit of large optical depth, all of the stellar luminosity is trapped and re-radiated, so that the momentum flux is everywhere the full $d\tau\,L/c$ directed radially from the clump center of density and follows the scalings we adopt here. 
This is trivially true in spherical or cylindrical (filamentary) geometries, but is a good approximation even for complex density distributions if the optical depths are sufficiently large.
This is an important distinction that makes radiation pressure a particularly efficient feedback mechanism 
in dense regions (relative to other sources of energy or momentum such as SNe or stellar winds).
In Appendix~\ref{sec:appendix:leakage} we discuss 
the more complicated case of an inhomogeneous density distribution. However, to the extent that it modifies 
our conclusions, it is usually equivalent to variations in the net efficiency (encapsulated in $\eta_{p}$), 
rather than the spatial distribution or direction of the flux. Of course, if desired, the momentum could be 
isolated to each star by simply taking the limit $N_{\rm sml}\rightarrow0$.\footnote{In 
Appendix~\ref{sec:appendix} we show that directing the momentum from the 
cloud density peak, center of gas mass, or center of stellar mass or luminosity makes no difference 
to our conclusions. Likewise allowing for more complex geometries by directing momentum 
along the local density gradients gives nearly identical results.}

Given that the local density structure of the gas is at least partially resolved, we use this 
information to estimate the IR optical depth $\tau_{\rm IR}  = {\small \Sigma}_{\rm eff}\,\kappa_{\rm IR}$
where ${\small \Sigma}_{\rm eff} \simeq M_{\rm clump}/\pi R_{\rm clump}^2$ is the gas surface density of the clump of interest.\footnote{For a log-normal density distribution within a given clump, the effective optical depth of the inhomogeneous clump is typically within 30\% of the mean optical depth \citep{murray:molcloud.disrupt.by.rad.pressure}.   Thus using the latter to determine $\dot p_j$ is sufficiently accurate for our purposes. See also \S~\ref{sec:results:global:structure}.}
The opacity at IR wavelengths is approximately constant for dust temperatures $\sim 100-1000$ K, so we adopt  $\kappa_{\rm IR}\approx 5\,{\rm cm^{2}\,g^{-1}}$ (this is convenient given that we are not performing radiative transfer and thus do not have information about the true dust temperature). Note that both the weighting of $L_{j}$ and this calculation of $\tau_{\rm IR}$ implicitly scale so that gas near the density center where the flux and optical depth are largest will be more strongly accelerated than gas in the outskirts of the system.

We can apply the force associated with $\dot p_j$ from equation~\ref{eqn:mom.vs.L} in two ways, either stochastically or as a continuous acceleration (the latter is the simplest to implement in grid-based calculations).   In the stochastic model, we model the momentum deposition by randomly ``kicking'' particles, with an average mass flux set by
\begin{equation}
\dot{M}_{w}\,v_{w} = \dot{p}_{j}
\end{equation}
where $\dot{M}_{w}$ is the mass-loading and $v_{w}$ is the initial velocity.
What is the appropriate ``initial'' velocity $v_w$?  Models of momentum-driven outflows 
argue that gas should be accelerated to the {\em local} escape velocity from 
the star clusters and/or clouds from which they are launched \citep{murray:molcloud.disrupt.by.rad.pressure}.  We therefore take 
\begin{equation}
v_{\rm w} \approx v_{\rm esc} \approx \eta_{v}\,v_{\rm dim}(M_{\rm clump},\,\rho,\,...)
\label{eq:v}
\end{equation}
where $v_{\rm dim}$ is an estimate of the 
escape velocity as a function of the simulation parameters and
$\eta_{v}$ is a normalization parameter that accounts for details such as 
the exact mass profile shape, microphysical acceleration as a function of 
position, etc.   In practice, we have experimented with a variety of choices for the  
velocity and will show that it makes relatively little difference.  This is because the key parameter that determines the effect of the feedback is the total momentum/force (eq. \ref{eqn:mom.vs.L}).

The escape velocity from 
the clump as resolved by our simulations is 
$v_{\rm dim} \approx \sqrt{2\,G\,M_{\rm clump}/R_{\rm clump}}$. 
However, some fraction of the clump will turn into stars
in a dense stellar cluster, the internal dynamics and peak density of which are unresolved. The true relevant escape velocity from the location where the outflows are driven is probably the escape velocity from that cluster. 
We therefore take the mass in young stars in the clump to 
be the ``star cluster'' mass and use the empirical size-mass 
relation of clusters (e.g., \citealt{murray:sizes.lum.star.clusters})  to determine the cluster escape velocity:   
\be
v_{\rm dim} =  \left(\frac{2\,G\,M_{\ast,\,{\rm cl}}}{R_{\rm cl}}\right)^{1/2}
\approx 66 \,\left(\frac{M_{\ast,\,{\rm cl}}}{10^{6}\,M_{\sun}}\right)^{1/4} \,{\rm km\,s^{-1}} \label{eqn:vdim}\ee
for $M_{\ast,\rm cl}\sim 10^{5}-10^{9}\,M_{\sun}$. 
In our models, we take $v_{\rm dim}$ to be the maximum of 
either the resolved clump escape velocity or the 
inferred star cluster escape velocity; the latter is almost always larger.
In a timestep $\Delta t$, the probability that the particle of mass $M_{\rm gas, \,j}$ is "kicked"  is then given by 
\begin{equation}
\label{eqn:prob.kick}
P_{\rm w} = 1-\exp{\{-(\dot{M}_{w}\,\Delta t)/M_{\rm gas, \,j}\}}.
\end{equation}
The particle then has a momentum $\Delta p_{j} = M_{\rm gas, \,j}\,{\bf v_{w}}$ added 
to its initial momentum, directed radially away from the clump center.

In addition to the stochastic acceleration of particles described above, we can alternatively accelerate the particles continuously rather than with individual ``kicks''; 
in this case the particle is simply given a $\Delta v_{j} = \dot{p}_{j}\,\Delta t/M_{\rm gas, \,j}$ 
every timestep. Which prescription is more physically appropriate depends on 
whether the outflows generated by stellar feedback are being accelerated at reasonably large radii (e.g.\ at the outskirts of  clouds), or whether they are launched in the dense central regions 
near the star cluster. 
We show below that the two methods yield similar results.

Many implementations of stellar feedback in the literature turn off the hydrodynamics, pressure 
forces, cooling, and/or star formation for some period of time, often chosen such that a wind escapes the galaxy entirely \citep[or until the wind reaches some distance from its launching point; see][]{springel:multiphase,
oppenheimer:metal.enrichment.momentum.winds,
sales:2010.cosmo.disks.w.fb}.   In our models, by contrast, there is no such modification of the underlying equations.   We are able to directly model the feedback and the resulting dissolution of star clusters for three reasons: 
first, our high resolution allows us to resolve a multi-phase ISM structure into 
which outflows can propagate; second, we identify massive star-forming regions and drive outflows coherently from them, rather than randomly within those regions;  and third, because the feedback is momentum-driven,  it drives strong turbulence even in dense, highly radiative environments.    In situations where a lower resolution is inevitable (e.g. cosmological simulations), it may be necessary for numerical reasons to modify the methods proposed here in order to maintain an efficient source of stellar feedback.  This will be studied in future work.

The feedback model used in this paper is ultimately defined by the 
two key parameters $\eta_{p}$ and $\eta_{v}$ (eqs. \ref{eqn:mom.vs.L} \& \ref{eq:v}).
We will discuss the consequences of different choices for these parameters below;  we take $\eta_p \sim 1$ and $\eta_v \sim 1$ as our physically-motivated default values. 

\begin{footnotesize}
\ctable[
  caption={{\normalsize Galaxy Models}\label{tbl:sim.ics}},center,star
  ]{lccccccccccccccc}{
\tnote[ ]{Parameters describing our galaxy models, used as the initial conditions for all of the simulations: \\ 
{\bf (1)} Model name: shorthand for models of a high-redshift massive 
starburst ({\bf HiZ}); local gas-rich galaxy ({\bf Sbc}); MW-analogue ({\bf MW}); and isolated SMC-mass dwarf ({\bf SMC}). 
{\bf (2)} $\epsilon_{g}$: gravitational force softening in our highest-resolution simulations (ultra-high-res).  ``High-res'' sims use twice this value.  ``Intermediate-res'' four times this value. 
{\bf (3)} $m_{i}$: Gas particle mass in our highest-resolution simulations (ultra-high-res).  ``High-res'' sims use eight times this value.  ``Intermediate-res'' fifty times this value. New star particles formed have mass $=0.5\,m_{i}$, disk/bulge particles $\approx m_{i}$, and dark matter halo particles $\approx5\,m_{i}$. 
{\bf (4)} $M_{\rm halo}$: halo mass. 
{\bf (5)} $c$: halo concentration. Values lie on the halo mass-concentration 
relation at each redshift ($z=0$ for SMC, Sbc, and MW; $z=2$ for HiZ). 
{\bf (6)} $V_{\rm max}$: halo maximum circular velocity. 
{\bf (7)} $M_{\rm bary}$: total baryonic mass.
{\bf (8)} $M_{\rm b}$: bulge mass.
{\bf (9)} $a$: \citet{hernquist:profile} profile scale-length for bulge. 
{\bf (10)} $M_{d}$: stellar disk mass.
{\bf (11)} $r_{d}$: stellar disk scale length. 
{\bf (12)} $h$: stellar disk scale-height.
{\bf (13)} $M_{g}$: gas disk mass.
{\bf (14)} $r_{g}$: gas disk scale length (gas scale-height determined so that $Q=1$).
{\bf (15)} $f_{\rm gas}$: average gas fraction of the disk 
inside of the stellar $R_{e}$ ($M_{\rm g}[<R_{e}]/(M_{\rm g}[<R_{e}]+M_{\rm d}[<R_{e}])$).
The total gas fraction, including the extended disk, is $\sim50\%$ larger.
{\bf (15)} $t_{\rm dyn}$: gas disk dynamical time at the half-gas mass radius. 
}
}{
\hline\hline
\multicolumn{1}{c}{Model} &
\multicolumn{1}{c}{$\epsilon_{\rm g}$} &
\multicolumn{1}{c}{$m_{i}$} &
\multicolumn{1}{c}{$M_{\rm halo}$} & 
\multicolumn{1}{c}{$c$} & 
\multicolumn{1}{c}{$V_{\rm max}$} & 
\multicolumn{1}{c}{$M_{\rm bary}$} & 
\multicolumn{1}{c}{$M_{\rm b}$} & 
\multicolumn{1}{c}{$a$} & 
\multicolumn{1}{c}{$M_{\rm d}$} & 
\multicolumn{1}{c}{$r_{d}$} & 
\multicolumn{1}{c}{$h$} & 
\multicolumn{1}{c}{$M_{\rm g}$} & 
\multicolumn{1}{c}{$r_{g}$} &
\multicolumn{1}{c}{$f_{\rm gas}$} &
\multicolumn{1}{c}{$t_{\rm dyn}$} \\
\multicolumn{1}{c}{\,} &
\multicolumn{1}{c}{[pc]} &
\multicolumn{1}{c}{[$\msun$]} & 
\multicolumn{1}{c}{[$\msun$]} & 
\multicolumn{1}{c}{\,} & 
\multicolumn{1}{c}{[${\rm km\,s^{-1}}$]} & 
\multicolumn{1}{c}{[$\msun$]} & 
\multicolumn{1}{c}{[$\msun$]} & 
\multicolumn{1}{c}{[kpc]} & 
\multicolumn{1}{c}{[$\msun$]} & 
\multicolumn{1}{c}{[kpc]} & 
\multicolumn{1}{c}{[pc]} & 
\multicolumn{1}{c}{[$\msun$]} & 
\multicolumn{1}{c}{[kpc]} &
\multicolumn{1}{c}{\,} &
\multicolumn{1}{c}{[Myr]} \\
\hline
{\bf Sbc} & 2.5 & 130 & 1.5e11 & 11 & 86 & 1.05e10 & 1e9 & 0.35 & 4e9 & 1.3 & 320 & 5.5e9 & 2.6 & 0.36 & 22 \\ 
{\bf HiZ} & 3.5 & 1700 & 1.4e12 & 3.5 & 230 & 1.07e11 & 7e9 & 1.2 & 3e10 & 1.6 & 130 & 7e10 & 3.2 & 0.49 & 12 \\ 
{\bf MW} & 2.5 & 220 & 1.6e12 & 12 & 190 & 7.13e10 & 1.5e10 & 1.0 & 4.73e10 & 3.0 & 300 & 0.9e10 & 6.0 & 0.09 & 31 \\ 
{\bf SMC} & 0.7 & 25 & 2.0e10 & 15 & 46 & 8.9e8 & 1e7 & 0.25 & 1.3e8 & 0.7 & 140 & 7.5e8 & 2.1 & 0.56 & 45 \\ 
\hline\hline\\
}
\end{footnotesize}

\begin{footnotesize}
\ctable[
  caption={{\normalsize Simulations Plotted in This Paper}\label{tbl:sims}},center
  ]{lcccc}{
\tnote[ ]{Parameters of our key simulations (only simulations appearing in Figures are listed; 
others are noted in the text):\\ 
 \  
{\bf (1)} Name/ID. First characters correspond to 
the class of galaxy model (``SMC,'' ``MW,'' ``Sbc,'' or ``HiZ,'' as in Table \ref{tbl:sim.ics}). \\
 \  
{\bf (2)} Total particle number. \\
 \  
{\bf (3)} Star formation law. ``--''  corresponds to the default law: 
$\dot{\rho}_{\ast}=\epsilon\,\rho/t_{\rm ff}(\rho)$ for $\rho > \rho_0$, with $\epsilon = 1.5\%$ and $n_0 = 100$ cm$^{-3}$; varied quantities are noted (see Fig.~\ref{fig:sfh.sflaw}). \\
 \  
{\bf (4)} Momentum-loading normalization $\eta_{p}$ (see eq.~\ref{eqn:mom.vs.L}). \\
 \  
{\bf (5)} Initial velocity normalization $\eta_{v}$ (see eq.~\ref{eq:v}).}
\tnote[a]{Acceleration is continuous rather than discrete ``kicks.''\\ 
}
}{
\hline\hline
\multicolumn{1}{c}{\ \ \ \ Simulation\ \ \ \ } &
\multicolumn{1}{c}{\ \ \ \ $N_{\rm part}$\ \ \ \ } &
\multicolumn{1}{c}{SF Law} & 
\multicolumn{1}{c}{\ \ \ \ $\eta_{p}$\ \ \ \ } & 
\multicolumn{1}{c}{\ \ \ \ $\eta_{v}$\ \ \ \ } \\
\hline
{\bf HiZ\_8\_0\_nofb} & 2e6 & -- & -- & -- \\ 
{\bf HiZ\_8\_2\_nofb} & 2e7 & -- & -- & -- \\ 
{\bf HiZ\_10\_4} & 2e6 & -- & 1.0 & 1.0 \\ 
{\bf HiZ\_8\_10} & 2e6 & -- & 2.0 & 2.0 \\ 
{\bf HiZ\_8\_11} & 2e6 & -- & 4.0 & 2.0 \\ 
{\bf HiZ\_9\_1} & 2e6 & -- & 10.0 & 1.0 \\ 
{\bf HiZ\_6\_0\_hr} & 1e7 &-- & 1.0 & 1.0 \\ 
{\bf HiZ\_7\_0\_hr} & 6e7 & -- & 1.0 & 1.0 \\ 
{\bf HiZ\_6\_3\_hr} & 1e7 & $\dot{\rho}\propto\rho^{1.0}$ & 1.0 & 1.0 \\ 
{\bf HiZ\_6\_4\_hr} & 1e7 & $\dot{\rho}\propto\rho^{2.0}$ & 1.0 & 1.0 \\ 
{\bf HiZ\_7\_1\_hr} & 6e7 & $\epsilon=0.35\%$ & 1.0 & 1.0 \\ 
{\bf HiZ\_7\_2\_hr} & 6e7 & $\epsilon=6.0\%$ & 1.0 & 1.0 \\ 
{\bf HiZ\_7\_3\_hr} & 6e7 & $n_{\rm c}=2500$ & 1.0 & 1.0 \\ 
{\bf HiZ\_7\_4\_hr} & 6e7 & $n_{\rm c}=25$ & 1.0 & 1.0 \\ 
{\bf HiZ\_10\_4\_hr} & 2e7 & -- & 1.0 & 1.0 \\ 
{\bf HiZ\_10\_5\_hr} & 2e7 & -- & 1.0 & 2.0 \\ 
{\bf HiZ\_10\_6\_hr} & 2e7 & -- & 1.0 & 4.0 \\ 
{\bf HiZ\_10\_7\_hr} & 2e7 & -- & 2.0 & 1.0 \\ 
{\bf HiZ\_10\_8\_hr} & 2e7 & -- & 4.0 & 1.0 \\ 
{\bf HiZ\_10\_9\_hr} & 2e7 & -- & 10.0 & 1.0 \\ 
{\bf HiZ\_10\_14\_hr} & 2e7 & -- & 0.33 & 1.0 \\ 
{\bf HiZ\_10\_11\_hr} & 2e7 & -- & 1.0 & --$^{a}$ \\ 
{\bf HiZ\_8\_14\_hr} & 2e7 & -- & 4.0 & 2.0 \\ 
{\bf HiZ\_8\_17\_hr} & 2e7 & -- & 5.0 & 4.0 \\ 
{\bf HiZ\_10\_4\_uhr} & 2e8 & -- & 1.0 & 1.0 \\ 
\hline\hline
{\bf MW\_8\_3\_nofb} & 3e6 & -- & -- & -- \\ 
{\bf MW\_9\_1} & 2e6 & -- & 1.0 & 1.0 \\ 
{\bf MW\_10\_7\_hr} & 1e7 & -- & 1.0 & 1.0 \\ 
{\bf MW\_10\_8\_hr} & 1e7 & $\epsilon=0.35\%$ & 1.0 & 1.0 \\ 
{\bf MW\_10\_9\_hr} & 1e7 & $\epsilon=6.0\%$ & 1.0 & 1.0 \\ 
{\bf MW\_10\_10\_hr} & 1e7 & $\dot{\rho}\propto\rho^{1.1}$ & 1.0 & 1.0 \\ 
{\bf MW\_10\_11\_hr} & 1e7 & $\dot{\rho}\propto\rho^{2.0}$ & 1.0 & 1.0 \\ 
{\bf MW\_10\_12\_hr} & 1e7 & $n_{\rm c}=10$ & 1.0 & 1.0 \\ 
{\bf MW\_10\_13\_hr} & 1e7 & $n_{\rm c}=1000$ & 1.0 & 1.0 \\ 
{\bf MW\_9\_1\_hr} & 2e7 & -- & 1.0 & 1.0 \\ 
{\bf MW\_9\_2\_hr} & 2e7 & -- & 1.0 & --$^{a}$ \\ 
{\bf MW\_9\_3\_hr} & 2e7 & -- & 1.0 & 2.0 \\ 
{\bf MW\_9\_4\_hr} & 2e7 & -- & 0.33 & 1.0 \\ 
{\bf MW\_9\_5\_hr} & 2e7 & -- & 10.0 & 1.0 \\ 
{\bf MW\_8\_4\_hr} & 3e7 & -- & 10.0 & 4.0 \\ 
{\bf MW\_8\_5\_hr} & 3e7 & -- & 10.0 & 1.0 \\ 
{\bf MW\_10\_2\_hr} & 3e7 & -- & 10.0 & 2.0 \\ 
{\bf MW\_10\_4\_hr} & 3e7 & -- & 1.0 & 1.0 \\ 
{\bf MW\_9\_1\_uhr} & 2e8 & -- & 1.0 & 1.0 \\ 
{\bf MW\_8\_uhr} & 3e8 & -- & 10.0 & 2.0 \\ 
\hline\hline
{\bf SMC\_10\_3\_nofb} & 2e7 &  -- & -- & -- \\ 
{\bf SMC\_10\_1\_hr} & 2e7 &  -- & 4.0 & 2.0 \\ 
{\bf SMC\_10\_2\_hr} & 2e7 &  -- & 10.0 & 2.0 \\ 
{\bf SMC\_10\_4\_hr} & 2e7 &  -- & 1.0 & 1.0 \\ 
{\bf SMC\_10\_uhr} & 1e9 &  -- & 10.0 & 2.0 \\ 
\hline\hline
{\bf Sbc\_10\_3\_nofb} & 2e7 &  -- & -- & -- \\ 
{\bf Sbc\_10\_1\_hr} & 2e7 &  -- & 4.0 & 2.0 \\ 
{\bf Sbc\_10\_2\_hr} & 2e7 &  -- & 10.0 & 2.0 \\ 
{\bf Sbc\_10\_4\_hr} & 2e7 &  -- & 1.0 & 1.0 \\ 
{\bf Sbc\_10\_uhr} & 2e8 &  -- & 10.0 & 2.0 \\ 
\hline\hline
}
\end{footnotesize}

\subsection{Galaxy Models}
\label{sec:ICs}

Our goal in this paper is to study the effects of stellar feedback on the ISM structure and star formation
in galaxies.   We do so using idealized models of disk galaxies with initial conditions motivated by galaxies in both the local and high-redshift Universe. We do not attempt to model the cosmological evolution of these 
disks, and so do not include extended gaseous halos or cold flows from which they would accrete. Rather, our goal is to study how a given feedback mechanism will change behavior given a specific set of (observed) disk properties.
The methodology for building the initial galaxies follows that described 
in detail in a series of papers 
\citep[see e.g.][]{springel:models,dimatteo:msigma,robertson:msigma.evolution,cox:kinematics,
  younger:minor.mergers}. 
The disks each include an extended dark matter halo with an NFW profile \citep{nfw:profile}, a stellar bulge 
(typically with a \citealt{hernquist:profile} profile), and exponential stellar and gaseous disks.   In all of the models, the initial vertical pressure support for the gas disk is provided by thermal pressure.   As we describe in \S \ref{sec:results:global} and \ref{sec:results:sfh}, however, this thermal energy is quickly radiated away and the system approaches a new statistical equilibrium with star formation and stellar feedback determining many of the properties of the gas disk.    

The simulations are carried out at several different resolutions: the ``standard'' resolution has a total of $\approx3\times10^{6}$ particles, with $\approx10^{6}$ particles in the gas+stellar disk (the initial bulges are small and so have fewer particles -- thus most of the remaining 
particles are in the dark matter halo). 
Our ``high'' resolution simulations use $10$ times as many particles, reaching $\approx 10^{7}$ particles in the disk.   We also have at least one "ultra-high" resolution simulation per galaxy model with $> 10^8$ particles in the disk (to our knowledge, these are the highest-resolution galaxy-scale SPH simulations that have been performed to date). 
The models are generally all run for $\approx20$ dynamical times at $R_{e}$ ($\approx3$ orbital times), 
but they typically converge to quasi-steady state behavior after just $\approx4-5\,t_{\rm dyn}$. After this the evolution is essentially just slow, steady-state gradual gas exhaustion; we have confirmed this in at least one run of each galaxy model run for $5$ times longer than the ``standard'' runs.
As described below, the spatial and mass resolutions in each of the simulations depend on the galaxy model.  

We consider four galaxy models, motivated by $z \sim 2$ high star formation rate galaxies (non major mergers), local low-luminosity LIRGs, Milky Way like spirals, and SMC-like dwarf galaxies.   The basic properties of these models are summarized in Table \ref{tbl:sim.ics}.

\noindent {\bf Sbc:} This simulation is designed to model an intermediate-mass, 
relatively gas-rich star-forming disk in the local universe (e.g.\, a low-luminosity LIRG with 
$L_{\rm bol}\sim10^{10-11}\,L_{\sun}$ and $\dot{M}\sim1-10\,\msunyr$). 
The galaxy has a total baryonic mass $1.05\times10^{10}\,\msun$, 
with a bulge having a mass $M_{b}=10^{9}\,\msun$ and a \citet{hernquist:profile} scale-length $a=350\,$pc; a stellar disk with a mass $M_{d}=4\times10^{9}\,\msun$ and an
exponential scale-length of $r_d = 1.3\,$kpc; and an extended gaseous disk with
$M_{g}=5.5\times10^{9}\,\msun$ and an exponential scale-length of $r_g = 2.6\,$kpc.  
The stellar disk has a ${\rm sech}^{2}$ vertical profile with a
scale height of $130\,$pc; it is initialized with a radial dispersion profile 
so that the local Toomre $Q=1$ at all positions. 
The gas disk is similarly initialized in vertical hydrostatic equilibrium with $Q=1$. 
The initial vertical support of the gas disk is via thermal pressure. 
The dark matter halo has a virial mass $M_{\rm halo}=1.5\times10^{11}\,\msun$, concentration $c=11$, and a spin parameter $\lambda=0.033$, chosen to match the 
typical concentrations and spins seen in cosmological simulations 
\citep{bullock:concentrations,vitvitska:spin}; this gives a total stellar-to-dark matter 
mass ratio similar to that inferred for systems of this mass 
\citep[e.g., by][]{moster:stellar.vs.halo.mass.to.z1}.   The disk is, however,  baryon-dominated 
within the central $\sim5-10\,$kpc, and as such may develop spiral and bar instabilities. 

For this galaxy model, our standard resolution has SPH smoothing lengths of $\sim5-10$\,pc in the central few kpc of the disk.  Our high resolution simulations have $\sim2-5\,$pc smoothing lengths and particle masses of $\sim1000\,\msun$, while the ultra-high resolution simulations  have particle masses of $100\,\msun$ and $1\,$pc resolution in the bulk of the disk.\footnote{The particular choice of gravitational softening is chosen as a compromise between 
matching the minimum SPH softening lengths, minimizing discreteness effects 
\citep[see e.g.][]{power:2003.nfw.models.convergence}, 
and giving a similar maximum resolvable density in each simulation ($\sim10^{5}\,{\rm cm^{-3}}$) 
that is much larger than the mean GMC density but still below densities where processes of individual star 
formation and detailed thermal physics become dominant.}

\noindent {\bf High-z:} This model is designed to approximate a massive, high-redshift, and 
strongly unstable disk forming stars at a very high rate $\sim100-400\,\msunyr$, typical of 
massive disks observed at $z\sim2-4$ \citep{erb:lbg.gasmasses, genzel:highz.rapid.secular,tacconi:high.molecular.gf.highz}. The galaxy has a baryonic mass 
$1.07\times10^{11}\,\msun$, with $M_{b}=7\times10^{9}\,\msun$ ($a=1.2$\,kpc), stellar disk $M_{d}=3\times10^{10}\,\msun$ ($r_d=1.6$\,kpc), gas disk $M_{g}=7\times10^{10}\,\msun$ ($r_g=3.2$\,kpc),  initialized with stellar scale-height $320\,$pc and $Q=1$ in gas and stars. 
This gives a typical gas fraction of $\sim0.5$ throughout the stellar and star-forming disk (with a larger 
HI gas reservoir at large radii). 
The halo has $M_{\rm halo}=1.44\times10^{12}\,\msun$ with $c = 3.5$ and a virial radius appropriate for that mass at $z=2$.   The system is baryon-dominated out to $\sim10\,$kpc.  The spatial and mass resolution in these simulations are somewhat larger than in the Sbc simulation because of the larger
total mass and spatial size of the disk; however, the Toomre mass and length-scale are also much larger, so this model is in a relative sense actually better resolved than the Sbc model. 

\noindent {\bf MW:} This system is initialized to represent a local, relatively gas-poor, Milky-Way like disk. The galaxy has a baryonic mass of $7.13\times10^{10}\,\msun$, a bulge with $M_{b}=1.5\times10^{10}\,\msun$, a stellar disk with $M_{d}=4.73\times10^{10}\,\msun$ ($r_d=3.0$\,kpc), and a gas disk with $M_{g}=0.9\times10^{10}\,\msun$ ($r_g=6.0$\,kpc).
The disk gas fraction is $f_g=0.05-0.10$ throughout the disk out to $\sim 8\,$kpc. 
The disk is initialized with a stellar scale-height $300\,$pc and $Q=1$.    
The halo has $M_{\rm halo}=1.6\times10^{12}\,\msun$, concentration $c=12$ and $R_{\rm vir}$ appropriate for $z=0$.   Observations suggest that the Milky Way hosts 
a pseudo-bulge or a bar instead of a classical bulge, so we initialize 
the bulge with a spherical exponential profile ($r_d=1.0\,$kpc), 
rather than a \citet{hernquist:profile} profile, but since the bulge mass 
is small this makes little difference to our conclusions.    
At our ultra-high (high) resolution, the force and mass resolution in the 
gas are $\approx2\,$pc (5\,pc) and $200\,\msun$ ($2000\,\msun$).

\noindent {\bf Dwarf/SMC:} This model is initialized to be similar to the inferred properties of the 
SMC \citep[before entering the MW halo, at least; see][and 
references therein]{besla:2010.mc.infall}, 
a typical low-mass, gas-rich dwarf. 
The galaxy has a baryonic mass 
$8.9\times10^{8}\,\msun$, with a
bulge having $M_{b}=10^{7}\,\msun$ ($a=0.25$\,kpc), 
a stellar disk with $M_{d}=1.3\times10^{8}\,\msun$ ($r_d=0.7$\,kpc), 
and gas disk with $M_{g}=7.5\times10^{8}\,\msun$ ($r_g=2.1$\,kpc).   The disk is
initialized with stellar scale-height $140\,$pc and $Q=1$. 
The halo has $M_{\rm halo}=2\times10^{10}\,\msun$, $c=15$ and $R_{\rm vir}$ appropriate for $z=0$.  
The system is dark-matter dominated at all radii outside of  the central few hundred pc.    For this model, our high resolution simulations have a spatial resolution and particle mass of $<1\,$pc and $\sim100\,\msun$, respectively.

\begin{figure*}
    \centering
    \scaleup
    \plotside{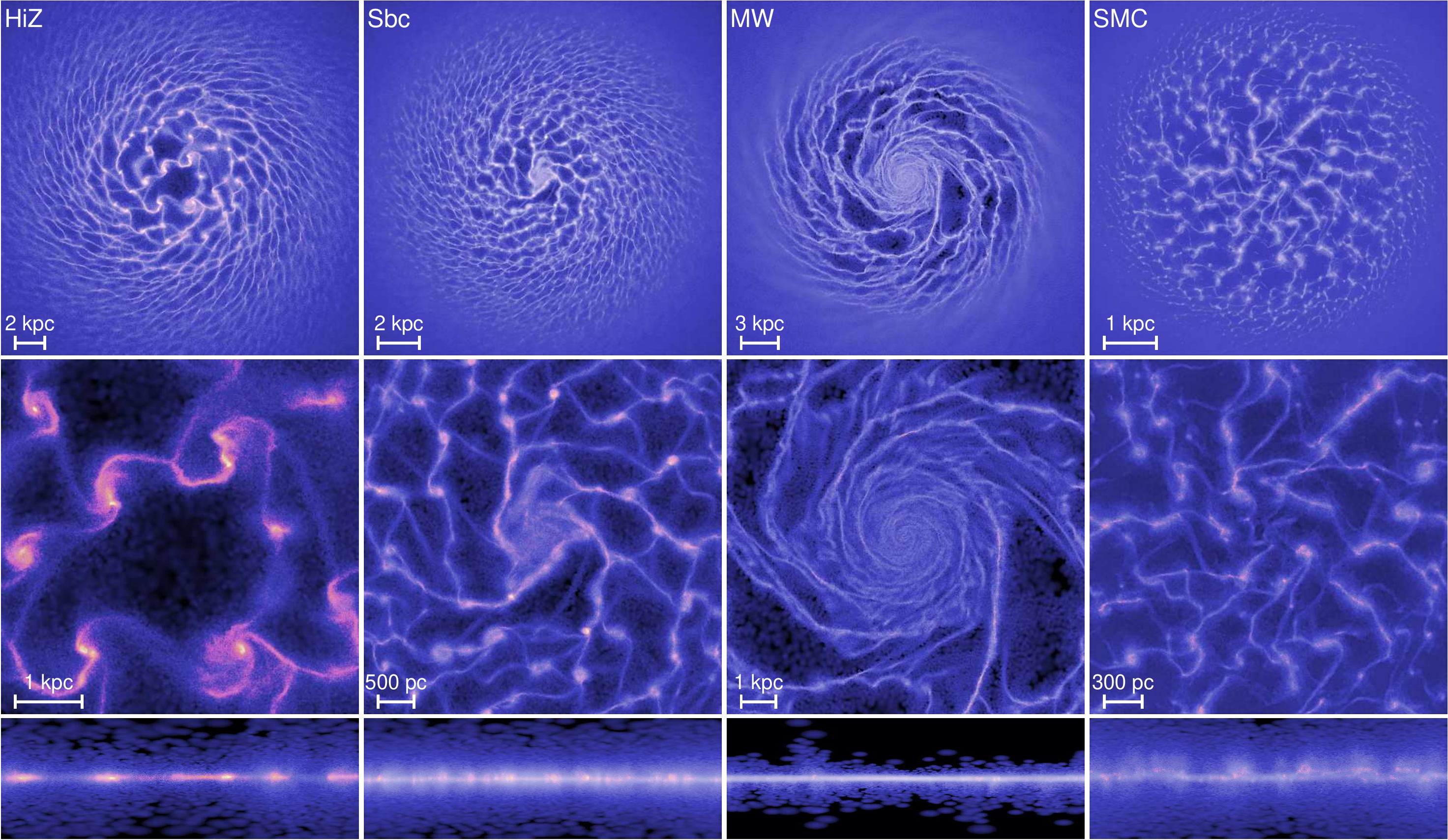}
    \caption{Images of the gas distribution for our fiducial simulations ($\eta_p = \eta_v = 1$)
    in the feedback-regulated quasi steady-state.    Brightness shows the gas surface density while color  shows the specific SFR (increasing from blue to red); both are on a logarithmic scale spanning a dynamic range of $\sim10^{6}$.   {\em Top:} Large scales (wide-field image) out to twice the half-gas mass radius.  
    {\em Middle:} Intermediate scales (zoom-in of the image at {\em top}) out to the half-SFR radius.   {\em Bottom:} Edge-on; scale is same as the middle image.     One example is shown for each of the initial conditions we 
    model (HiZ\_10\_4\_hr, Sbc\_10\_4\_hr, MW\_10\_4\_hr, and SMC\_10\_4\_hr in Table~\ref{tbl:sims}).
    The simulations develop complex substructure and exhibit a diverse range of gas morphologies.  Most stars are formed in dense but resolved giant `molecular' cloud complexes,  which are the sites of the feedback modeled here. 
       \label{fig:pics}}
\end{figure*}

\begin{figure}
    \centering
    \scaleup
    \plotone{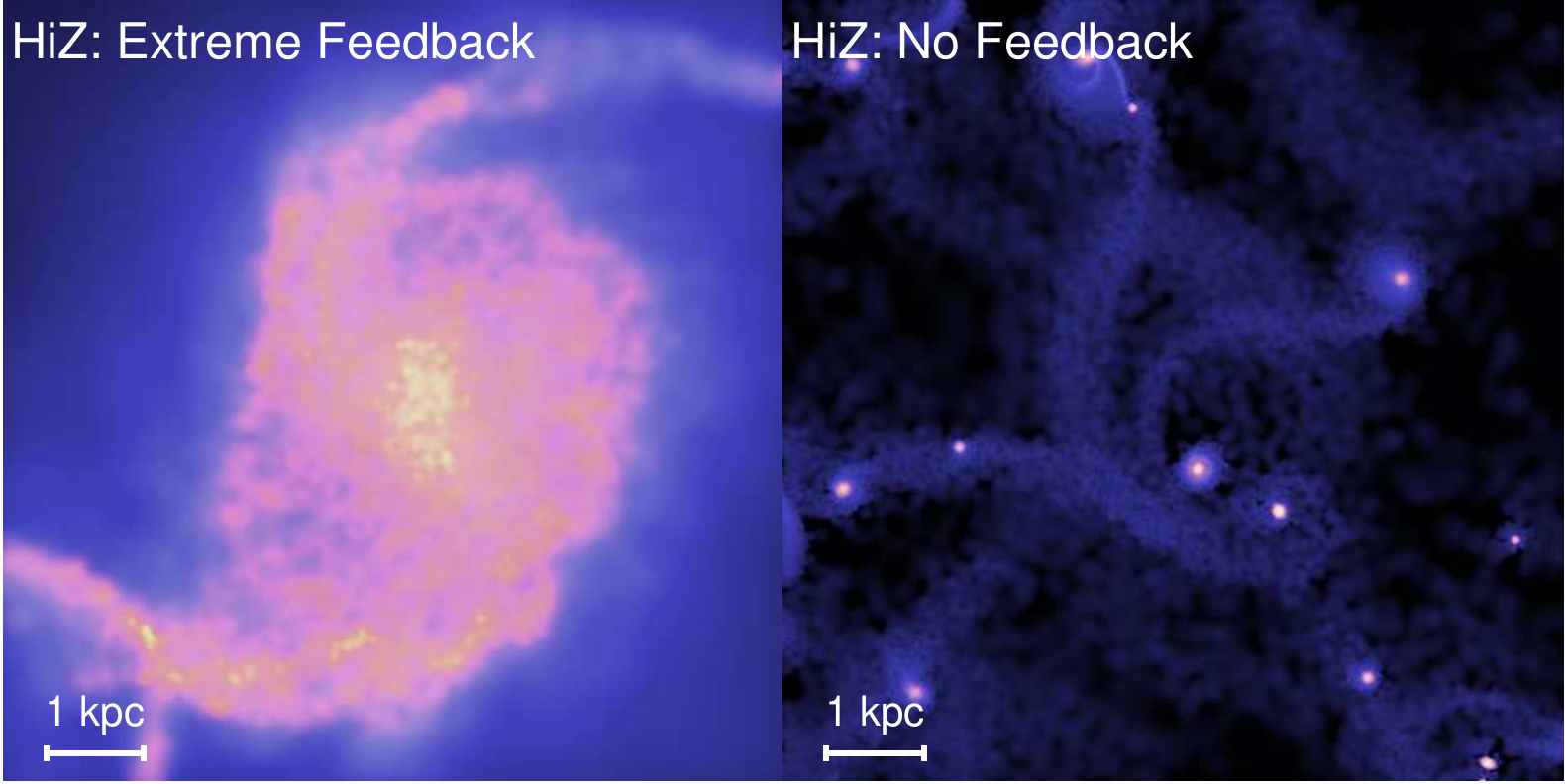}
    \caption{As Figure~\ref{fig:pics} ({\em middle left}), but for an otherwise identical HiZ simulation with 
    extremely strong feedback ({\em left}) with $\eta_{p}=100$ (this is not a realistic choice but purely shown for illustrative 
    purposes), and one with no feedback ({\em right}). 
    With arbitrarily strong feedback, all collapse of gas into GMC complexes is suppressed.
    With no feedback, the cloud complexes in Figure~\ref{fig:pics} undergo runaway collapse to the resolution limit 
    (the single white pixels at {\em right}); the mass piles up at densities $\gtrsim 10^{4}$ times larger than 
    in our ``standard'' models. 
       \label{fig:pics.vs.fb}}
\end{figure}

\section{Global Galaxy Properties}
\label{sec:results:global}

The key simulations described in this paper are summarized in Tables~\ref{tbl:sim.ics} and \ref{tbl:sims}. Table~\ref{tbl:sim.ics}  summarizes the properties of each galaxy model.   Table \ref{tbl:sims} summarizes the resolution of each simulation, the parameters of the star formation model, and the key feedback parameters $\eta_{p}$ and $\eta_{v}$ (eqs. \ref{eqn:mom.vs.L} \& \ref{eq:v}).

Figure~\ref{fig:pics} shows face-on and edge-on images of the gas surface density distribution for simulations of each galaxy model with our fiducial parameter choices $\eta_p = \eta_v = 1$.   Each image is shown in the quasi-steady feedback-regulated phase that sets in after a few dynamical times. The overall qualitative evolution is similar in all of the simulations with feedback.   The gas cools efficiently to low temperatures and collapses by gravitational instability at the Jeans/Toomre scale.   This leads to the formation of dense clumps that are the sites of star formation and, in our model, feedback.   The resolved density contrast between the centers of star-forming clumps and the interclump medium is typically $\sim 1000$ but can be as high as $\sim10^{6}$.   The ISM sustains this clumpy structure as long as we evolve our simulations, as gas is blown out of individual clumps (by feedback) into the more diffuse ISM before being incorporated into new dense clouds. We defer a rigorous analysis of the lifetimes and evolution of individual clumps for future work (in preparation) analyzing the properties of GMCs, where we can make rigorous comparisons with observations. But typically, we find average lifetimes of individual clouds $\lesssim10\,$Myr or a few free-fall times (weakly increasing with 
mass scale from the SMC through HiZ models), giving an integrated fraction $\sim1-5\%$ of clump mass turned into stars. 

This fragmentation is the natural extension of Jeans-mass GMCs in the MW and other nearby galaxies. Indeed, if we wish to explicitly resolve these scales, most of the gas mass {\em should} be in dense sub-clumps at something like the Jeans scale. The primary role of feedback is to regulate against runaway collapse and star formation in those clouds. 
Figure~\ref{fig:pics.vs.fb} illustrates how these morphologies depend on the strength of feedback. We consider the HiZ case, which is most strongly unstable, at two different extremes (holding all details of the model fixed, except feedback strength). First, with feedback much stronger than is realistic, $\eta_{p}=100$. In this case, essentially all sub-structure in the galaxy is ``wiped out,'' and the star formation is smoothly distributed over a $\sim10\,$kpc disk (despite $<10\,$pc resolution). This would be analogous to a MW model with no GMCs, where all star formation occurred in regions with local densities $\sim1\,{\rm cm^{-3}}$. Second, we consider a case with no feedback. In this extreme, the opposite occurs: the GMC complexes seen in Figure~\ref{fig:pics.vs.fb} dissipate their internal velocity dispersions and experience runaway collapse and star formation. This collapse proceeds until the GMCs reach the simulation resolution limit and leads to all of the gas being at extremely high densities, $n\sim10^{6}\,{\rm cm^{-3}}$ (as we show explicitly below);  a corollary is that the gas is converted into stars on essentially one (large-scale) dynamical time.

\begin{figure}
    \centering
    \plotone{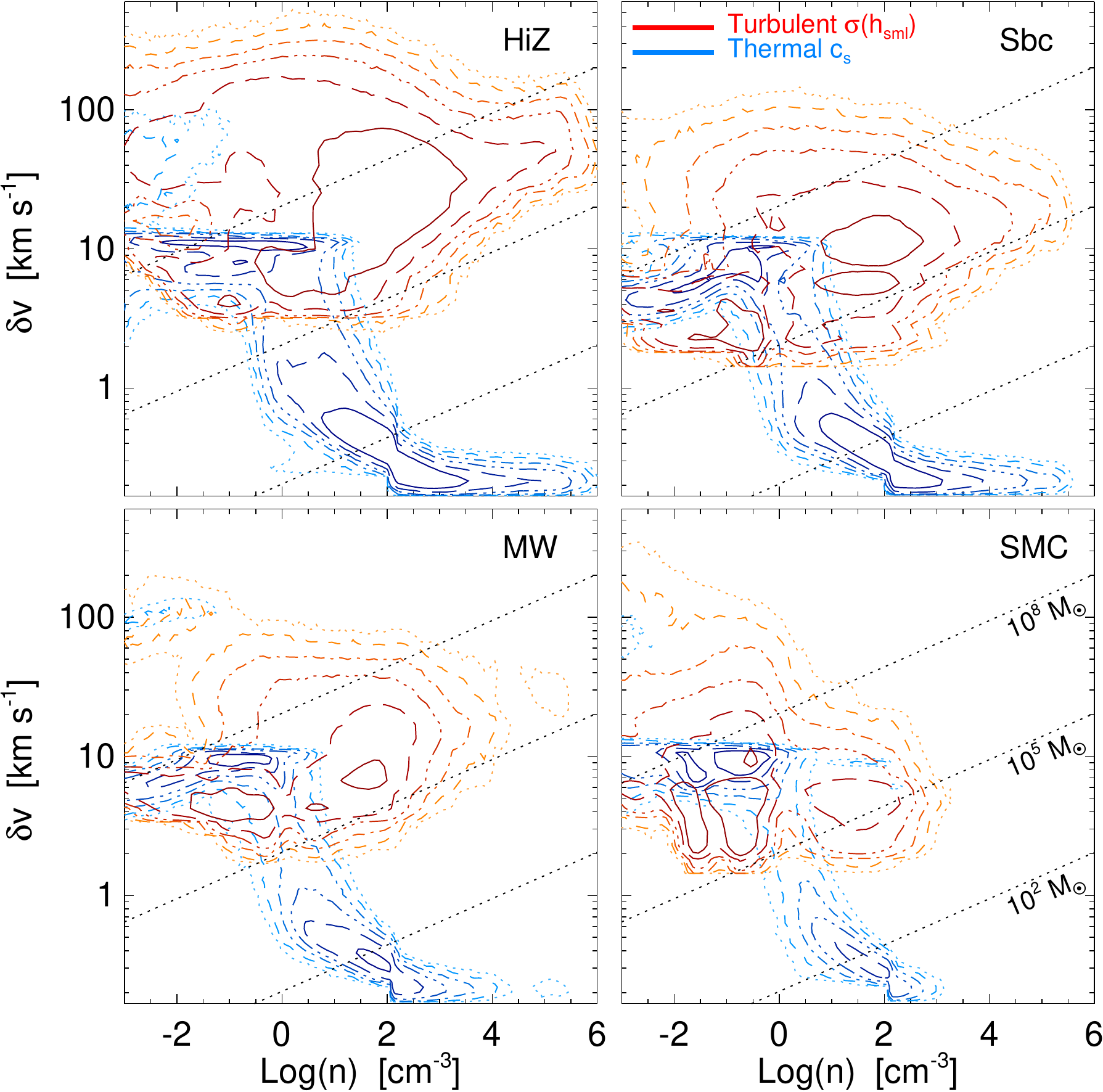}
    \caption{Phase diagram for the gas in the fiducial simulations in Figure~\ref{fig:pics}, at times in the feedback regulated quasi steady-state. Contours are iso-density at $\sim10^{-3,\,-2.5,\,-2,\,-1.5,\,-1,\,-0.5}$ of maximum (progressively darker dotted, short-dash, dot-dash, dot-dot-dot-dash, long-dash, solid contours, respectively). Blue contours show the thermal sound speeds $c_{s}\propto T^{1/2}$; red contours the local turbulent velocity dispersion $\sigma$ (averaged within one gas smoothing length $h_{\rm sml}$ around each particle). Lines of constant Jeans mass $\propto \delta v^{3}\,n^{-1/2}$ (black dotted) are shown for comparison. The median clump/cloud gas density is evident in the peak near $\sim100\,{\rm cm^{-3}}$. For all the dense gas, the thermal pressure is negligible compared to the turbulent pressure/velocities.     
As a result, the turbulent Jeans mass governs large-scale collapse and corresponds to the mass of massive clumps/GMCs (from $\sim10^{5}\,\msun$ in the SMC model through $\sim10^{8}\,\msun$ in the HiZ model); these are very well-resolved. The {\em thermal} Jeans masses are much smaller, but are only relevant for the dynamics on scales below the sonic length ($<0.1\,$pc) where individual groups of stars form; this is unresolved, hence the necessity of an ``effective'' small-scale star formation law. The slight ``upturn'' in $\sigma(h_{\rm sml})$ at $n\gg10^{4}\,{\rm cm^{-3}}$ (most evident in the HiZ model) comes from the minimum pressure corresponding to the \citet{truelove:1997.jeans.condition} Jeans condition (eq.~\ref{eqn:Pjeans}). This indicates where resolution limits prevent us from following further collapse to higher densities. 
    \label{fig:phase.diagram}}
\end{figure}

Figure~\ref{fig:phase.diagram} shows the phase diagram for the gas in each of our fiducial simulations: we plot both the thermal sound speeds and turbulent velocities as a function of gas density (averaged over the smallest available scale, the SPH smoothing length).   For low density gas with $n\ll 1\,{\rm cm^{-3}}$, the sound speed and turbulent velocity are often comparable, but for denser gas the turbulent velocity is always much larger than the thermal sound speed. 

The characteristic densities of clumps/GMCs are evident in the peak of the gas distributions near $n\sim100\,{\rm cm^{-3}}$ in Figure~\ref{fig:phase.diagram};  the typical turbulent mach numbers for this gas are $\sim30-100$.   
Because of the high Mach numbers, turbulent motions rather than thermal motions are the dominant impediment to gravitational collapse.  Specifically, the characteristic mass of large GMCs is set by the turbulent Jeans mass for the bulk of the matter, and corresponds to: $\sim10^{5}\,\msun$ in the SMC case, $\sim10^{6}\,\msun$ in the MW and Sbc cases, and $\sim10^{8}\,\msun$ in the HiZ case.   These estimates agree reasonably well with the observed properties of massive cloud complexes in the respective systems.  By contrast, if the gas were thermally supported, the characteristic mass of collapsed gas would be much smaller.   For the dense gas, however, thermal support is only important on scales below the sonic length ($\lesssim0.1\,$pc), which is well below our resolution limit.

The minimum pressure to prevent unresolved collapse below the resolution limit (eq.~\ref{eqn:Pjeans}) is well below the resolved turbulent pressure for the median densities in Figure~\ref{fig:phase.diagram}. This effective pressure does, however, produce the small ``upturn'' in the turbulent $\delta v$ at the very highest densities $n\gg10^{4}\,{\rm cm^{-3}}$. For our purposes, the key point is that we resolve the median GMC length, density, and mass scales well, even in our lowest-resolution models. 

\subsection{Morphologies}
\label{sec:results:global:morph}

There are a variety of morphologies present in the simulated galaxies depending on how self-gravitating the disk is.   The high-redshift disk analogues (HiZ) are the most strongly self-gravitating and so  fragment into very massive clumps ($M_{\rm Toomre}\sim10^{8}-10^{9}\,\msun$), which dominate the star formation.   This morphology resembles the clumpy systems observed at $z \sim 2-3$ \citep{genzel:highz.rapid.secular,tacconi:smg.maximal.sb.sizes,law:dispersion.in.central.kpc.z2.disks}. 
This is even more clear when we focus on the region which contains half the star formation ({\em middle panel}) -- this is dominated by a few giant complexes. Viewed edge-on, the HiZ model appears qualitatively similar to the ``clump chain/cluster'' systems observed at high redshift.

The Sbc model fragments in a manner similar to that of the HiZ model.   However,  the disk is thinner, and the Jeans mass and length-scales are significantly smaller, so star formation is 
more distributed in many clouds (the number of massive clouds predicted at $Q \sim 1$ is $\sim (R/h)^2$ and is thus larger for thinner disks). At slightly later times than that shown 
in Figure~\ref{fig:pics}, the system develops a strong stellar bar, and the gas -- while still 
very clumpy -- flows into the center along the $m=2$ mode.   Flocculent spiral structure also develops in some of the Sbc runs at large radii.

\begin{figure*}
    \centering
    \plotsidesmallest{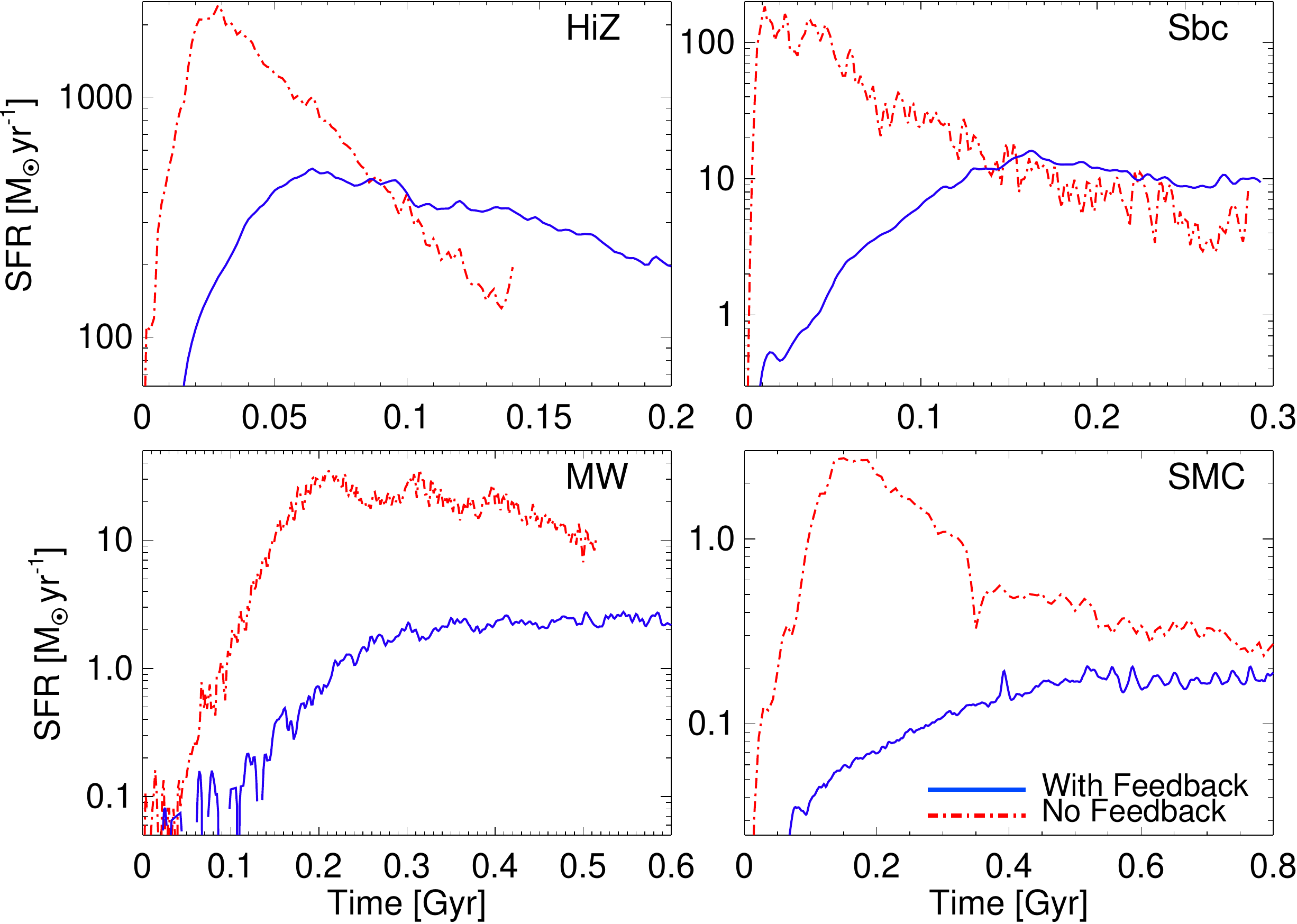}
    \caption{Total star formation rate for each of our galaxy models in Table \ref{tbl:sim.ics} as a function of time, both with feedback ($\eta_p = \eta_v = 1$) and without.  The timescales are different in each model and correspond to the characteristic dynamical timescales in each system (longer in the 
    more stable, dark-matter dominated systems; see Table \ref{tbl:sim.ics}).
    Absent feedback (red dot-dashed line) the gas collapses on a dynamical time, leading to a SFR well in excess of that observed in similar systems; the SFR then declines as the gas is exhausted.     With stellar feedback, the SFR reaches an approximate equilibrium in which feedback maintains 
    marginal stability to gravitational collapse ($Q \sim 1$).    \label{fig:sfh.ov}}
\end{figure*}

The Milky Way model in Figure \ref{fig:pics} shows clear grand design spiral structure.  It is also 
weakly barred in the center, though the bar feature is much more prominent 
in the stars. Note that this model has a lower gas fraction and is significantly more stable than the Sbc and HiZ models -- the latter because it is dark matter or bulge dominated at all radii.   As a result, the characteristic clump mass  is smaller ($M_{\rm Toomre}\sim10^{6}\,\msun$) and the Jeans length is much smaller relative to the effective radius 
(for a $Q\sim1$ disk, $\lambda_{\rm Jeans}\sim f_{\rm gas}^{2}\,R_{e}$, so $\sim100$ times smaller here). 
As a result, the individual ``clumps'' are much less prominent in the image, despite the fact that most of the mass in the star-forming disk does lie in thousands of resolved ``clouds'' with masses $\sim10^{4}-10^{6}\,\msun$. 
The small gas fraction also causes the disk to be significantly thinner in the edge-on image:  $Q \sim 1$ implies $h \sim f_g R$ for weakly self-gravitating disks (e.g., \citealt{thompson:rad.pressure}). In future work (in preparation), we will investigate the detailed structural properties of the ISM and simulated GMC-analogues to compare them to observations of the MW and local group galaxies. 

The SMC-like model behaves quite differently from the Milky Way model, although both are dark-matter dominated.  The SMC model is completely stable to global instabilities and thus forms stars in a more uniformly distributed fashion.   The ISM on these scales is turbulent and patchy, with an irregular or 
(on large scales) featureless structure, typical of observed dwarf galaxies. 
Despite the low SFR of $\sim0.1\,M_{\sun}\,{\rm yr^{-1}}$, the turbulent velocities 
generated by stellar feedback are sufficient to make the 
system quite ``puffy'' and thick (given the weaker potential depth). Figure \ref{fig:pics} shows that individual star-forming regions are resolved with size scales of $<10\,$pc. 

Note that because the gas in this model is quite low-density, the cooling times are long and energy input via supernovae and stellar winds will have a significant effect on the gas morphology. 
There are plain indications here that the present model, including momentum from radiation pressure alone, is not a complete description of the ISM. 
For example, the temperature of the ``diffuse'' ISM in {\em all} the galaxy models tends to be much too low. 
We show this explicitly in Figure~\ref{fig:phase.diagram}, where we plot 
the phase distribution of the gas.
The volume-filling gas distributed between dense clouds is almost entirely ``warm'' ($10^{4}\lesssim T \lesssim 10^{5}\,$K), with negligible mass in the characteristic ``hot phase'' of the ISM at $T\gtrsim10^{6}\,$K (there is some, generated by shocks, in the stronger HiZ and Sbc cases, but even here it is less than a percent of the total gas mass). 
Some additional heating mechanisms, such as SNe and ``fast'' stellar winds, are probably critical to explain the full temperature structure of the ISM. 
In future work we will investigate this in detail, with explicit models for various heating terms; for now, we simply note that the small mass fraction in the ``hot'' phase, while potentially important for phenomena such as galactic winds, is unlikely to change the structure of cold regions as it contains little mass and, even in MW-like galaxies, contributes only $\sim10\%$ to the typical ISM pressure \citep{boulares90}. We see in Figure~\ref{fig:phase.diagram} that the turbulent velocities are much larger in all dense gas than the thermal sound speeds (and tend to be near-virial), making the detailed thermal structure sub-dominant on these scales.

\subsection{Star Formation Histories}
\label{sec:results:sfh:fb}

Figure~\ref{fig:sfh.ov} shows the star formation history (galaxy-integrated star formation rate [SFR] as a function of time) of each of our galaxy models for the same feedback parameters used in Figure~\ref{fig:pics}; we also compare to simulations of the same galaxy models that include cooling and star formation, but not stellar feedback.  

In the models without feedback, the SFR increases to a peak value on a single global dynamical time; the SFR remains at this value until the gas in the disk is exhausted.  The peak SFRs in the simulations without feedback are a factor of $\gtrsim 10$ larger than those observed in the systems that motivate these galaxy models  -- the observed values are $\sim (50-300,\,3-20,\,2-4,\,0.1-0.5)\,\msun\,{\rm yr^{-1}}$ for high-z non-merging SMGs \citep{forsterschreiber:z2.sf.gal.spectroscopy}, low-z non-merging LIRGs \citep{sanders96:ulirgs.mergers, evans:ulirgs.are.mergers}, the MW and similar-mass 
spirals at $z=0$, and isolated SMC-mass systems at $z=0$ \citep{noeske:sfh,salim:2007.uv.ssfrs}.    In \S \ref{sec:results:ks} we explicitly show that these models also lie well off of the observed Kennicutt-Schmidt relation between SFR and gas surface density.   Physically, this is because in all of our simulations the gas can cool to an arbitrarily low temperature on a timescale short compared to the local dynamical time.  In the absence of feedback, the gas is then violently unstable to runaway clumping and rapid star formation.  The net result is that $\dot M_* \sim M_{\rm gas}/t_{\rm dyn}$, i.e., most of the gas is converted into stars on a single dynamical time, the timescale for the initially thermally supported gas disk to collapse. This behavior is physically correct in the absence of stellar feedback, and should be recovered in any simulation that does not include such feedback. 

A number of authors have suggested that instabilities due to self-gravity alone might generate the turbulence needed to slow down star formation in galaxies \citep[e.g.][]{ballesteros:gmc.core.lifetimes,tasker:2009.gmc.form.evol.gravalone,
krumholz:2010.instab.turb.in.disks}. 
Figure~\ref{fig:sfh.ov} is not consistent with this hypothesis.   Absent stellar feedback, the majority of the gas accumulates into dense clumps in which star formation proceeds unimpeded.  Independent simulations at similar resolution but with different physics included have reached the same conclusion \citep[e.g.][]{bournaud:2010.grav.turbulence.lmc}.  We thus find that stellar feedback is critical to regulating star formation in galaxies. A more subtle question is, when strong feedback is present, does it ``drive'' the turbulence, or is it still primarily driven by gravity? We will investigate this more quantitatively in future work. It generally appears, however, that the role of feedback is to offset the dissipation of turbulence and relative motions (particularly in dense regions), so in this sense it ``provides'' momentum; but the level it must provide, and the turbulent cascade and regulation of those motions, is primarily dictated by gravity. 

In contrast to the models without feedback, our simulations with stellar feedback rapidly reach a maximum SFR and then remain at this quasi-steady state for many dynamical times.    In some of our simulations, there is a slow increase in the SFR on a timescale longer than the disk's dynamical time; this is a consequence of both slowly-growing secular instabilities (e.g.\ halo bars) and spatial redistribution of gas by the stellar feedback (e.g.\ gas being driven in fountains from small to large radii). 
Eventually, however, since these are isolated systems without continuous gas accretion, the 
SFR must decline by gas exhaustion, but this decline is much more gradual than in the absence of feedback.  Test runs, run for $\sim5$ times longer, confirm that there is no new behavior after the first few dynamical times; the SFRs gradually decline as gas is exhausted. In \S \ref{sec:results:ks} we show that our simulations with stellar feedback are reasonably consistent with the the observed Kennicutt-Schmidt relation.   This is true for a range of feedback parameters.  Below we discuss the physical origin of this low star formation efficiency.


\begin{figure*}
    \centering
    \scaleup
    \plotside{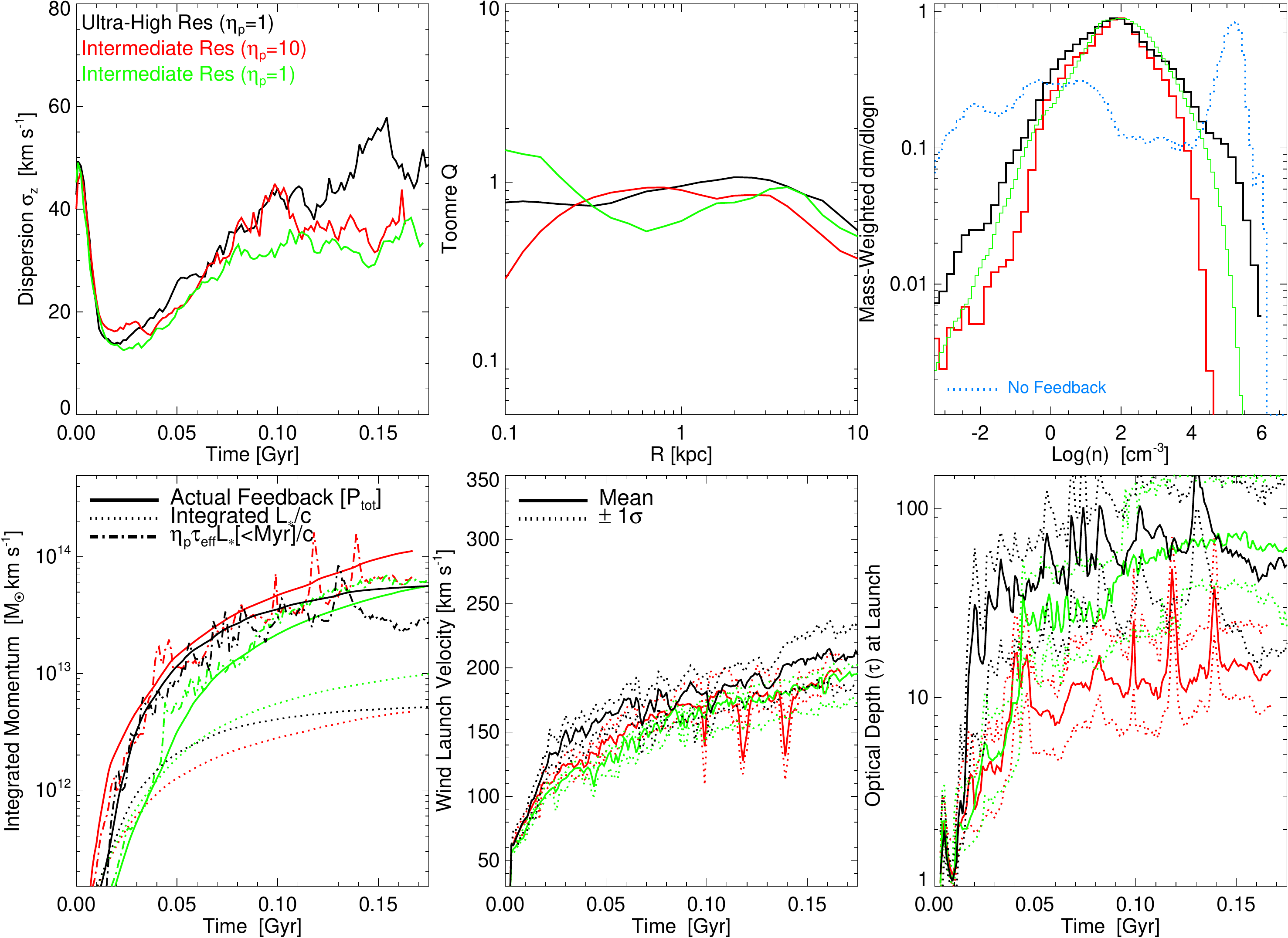}
    \caption{Properties of the ISM and feedback in several of our HiZ simulations:  intermediate (HiZ\_10\_4) and ultra-high resolution (HiZ\_10\_4\_uhr) simulations  with $\eta_{p}=1$ and $\eta_{v}=1$,
    and an intermediate-resolution (HiZ\_9\_1) simulation with  $\eta_{p}=10$ (see Table~\ref{tbl:sims}). 
    {\em Top Left:} Vertical gas velocity dispersion, $\sigma_{z} = \sqrt{c_s^2 + \delta v^2}$ 
    (averaged over the entire disk, weighted by gas mass).     The initial disk is thermally supported, but this thermally energy is rapidly radiated away; at later times  a comparable $\sigma_z$ is produced by feedback-driven turbulence.  {\em Top Center:}  Gas Toomre $Q$ parameter in narrow radial annuli as a function of radius (averaged over times $>60\,$Myr, when the system is quasi-steady state). 
    {\em Top Right:} Gas density distribution (gas mass per interval in log $n$) is lognormal with $\sim1-1.5$\,dex dispersion; low and high-res simulations converge to the same {\em median} density, but at low resolution the full width is not resolved.  With no feedback (dotted), the gas piles up at the highest resolvable densities.  {\em Bottom Left:} Sum of all momentum ($| \Delta p |$) injected via feedback (solid; eq.~\ref{eqn:mom.vs.L}) compared with input optical-UV stellar photon momentum $=\int L_{\ast}\,c^{-1}\,dt$ (dotted).   Note that the momentum injected is nearly the same for all three simulations, including $\eta_p = 1$ and $\eta_p = 10$.   The dot-dashed line shows that the input momentum is well-reproduced using the optical depths from the bottom right panel and only the very young stars ($<10^{6}$\,yr old).   This demonstrates that star-forming clusters disrupt rapidly.    {\em Bottom Center:} Mean "kick" velocity given to gas particles at their launching from 
    young stellar clusters (and 1-$\sigma$ dispersion); values approach $\sim150-200\,{\rm km\,s^{-1}}$, as expected given the massive $10^{8}\,\msun$ clumps  forming in these simulations.  The kick velocity is much larger than the actual dispersion in the disk because the particles shock and share their momentum immediately. 
    {\em Bottom Right:} Resolved IR optical depths of gas clumps used in the feedback model (eq. \ref{eqn:mom.vs.L}). 
    In the simulations with $\eta_p = 1$, $\tau \sim 30-50$, corresponding to $\Sigma \sim 10$ g cm$^{-2}$, comparable to the observed surface densities of star clusters on $\sim$pc scales. 
    The simulation with $\eta_p = 10$ has the same total input momentum ({\em bottom left} panel) but as a result the gas clumps only collapse to $\tau \sim 10$.
    A comparison of our MW-like models gives identical qualitative conclusions, but with 
    systematically shifted absolute values: $\sigma_{z}\sim10\,{\rm km\,s^{-1}}$, $Q\approx1$,  
    $\langle n \rangle\sim1\,{\rm cm^{-3}}$, ``kick'' $v\sim30-50\,{\rm km\,s^{-1}}$, and 
    $\langle \tau \rangle \sim 10-30$ at $\eta_{p}=1$.
        \label{fig:turb.highz.demo}}
\end{figure*}

\subsection{Structural Properties}
\label{sec:results:global:structure}

Figure~\ref{fig:turb.highz.demo} shows a number of the properties of the ISM in our HiZ model as a function of time and radius:  the vertical velocity dispersion ($\sigma_z$:  thermal and turbulent), the Toomre Q parameter\footnote{We define Q here as $\sigma\,\kappa/\pi\,G\,\Sigma_{\rm gas}$, where 
$\sigma$ is the full gas velocity dispersion, $\kappa$ is measured from the azimuthally-averaged mass profile, and 
$\Sigma_{\rm gas}$ is the gas surface density.} of the gas, the mass weighted density distribution function, the total momentum supplied to the gas, the "kick" velocity particles initially receive, and the (momentum-weighted) optical depth of gas clumps where the kicks are applied (i.e., the optical depth for the regions where most of the feedback occurs).    These results provide a more quantitative view of the quasi-steady feedback-regulated state reached in our simulations. 
We show the results for the HiZ model because the strong gravitational instability, high gas fraction, and very high star formation rate make it the model most sensitive to variations in our feedback prescription and give the largest differences between models with and without feedback. However, we find identical qualitative conclusions (discussed below), modulo the absolute value of the various quantities, for each of our other galaxy models. We focus on three different simulations in Figure~\ref{fig:turb.highz.demo}.   The first is one of our ultra-high resolution runs ({\em HiZ\_10\_4\_uhr}) with $\eta_p = \eta_v = 1$ and $2\times10^{8}$ particles, in which a typical Jeans-mass clump in the disk is resolved with as many as $\sim 10^{5}$ particles.   We compare this to a lower resolution simulation with the same feedback parameters ({\em HiZ\_10\_4}) and to a lower resolution simulation which has $\eta_{p}=10$ ({\em HiZ\_9\_1}) to compensate for the poorer resolution of the densest star-forming regions.   

Perhaps the most important result in Figure~\ref{fig:turb.highz.demo} is that the ISM properties do not depend sensitively on either resolution or the momentum feedback parameter $\eta_p$ (the star formation history does depend mildly on $\eta_p$ as we show in \S \ref{sec:results:sfh:model}).   The key reason for this is that the disk always self-regulates to maintain 
\begin{equation}
\label{eqn:Q}
Q \simeq \frac{\delta v\,\Omega}{\pi G\,\Sigma_g} \sim 1,
\end{equation}
where $\delta v$ is the turbulent velocity dispersion induced by the stellar feedback.  Figure~\ref{fig:turb.highz.demo} shows explicitly that all of the simulations maintain Q $\sim 1$ in the feedback-regulated phase  ({\em top middle panel}). 
The differences between models are small and all within the range of random variations and noise. Figure \ref{fig:turb.highz.demo} also shows the (mass-weighted) vertical velocity dispersion $\sigma_{z} = \sqrt{c_s^2 + \delta v_z^2}$ of the gas as a function of time ({\em top left panel}).  Initially $\sigma_z$ decreases rapidly as the thermal support is radiated away.  As star formation commences, however, stellar feedback quickly drives the turbulent velocity to $\delta v_z \sim 30-50\,{\rm km\,s^{-1}}$.   Given this turbulent velocity, the vertical scale-height of the disk is a few hundred pc, with only a modest dependence on radius; at all radii this thickness is much larger than the resolution limit.   

The early-time and late-time values of $\sigma_z$ in Figure~\ref{fig:turb.highz.demo} are comparable because in both limits $Q \sim 1$.   The models are initialized with thermal support and $Q = 1$ but this is quickly replaced by turbulent support that self-consistently maintains $Q \sim 1$ at later times.    The velocity dispersions in Figure~\ref{fig:turb.highz.demo} are also in reasonable  agreement with the observed values in high-redshift disks \citep{forsterschreiber:z2.disk.turbulence}.  
The other galaxy models also self-regulate at $Q\approx1$. However, given their lower masses, gas fractions, and 
star formation rates, this translates to lower absolute velocity dispersions: $\delta v \approx10\,{\rm km\,s^{-1}}$ in the MW and Sbc models, and $\approx6\,{\rm km\,s^{-1}}$ in the SMC model (modulo rescaling by this absolute value, however, the dependence of $\sigma_{z}$ on time, resolution, and $\eta_{p}$ is nearly identical to that shown in Figure~\ref{fig:turb.highz.demo}).

The {\em top right panel} in Figure~\ref{fig:turb.highz.demo} shows the mass-weighted gas density distribution averaged over the entire galaxy once the star formation reaches an approximate steady-state (since most of the gas mass is near $\sim 3$ kpc, the density distribution in an annulus at this radius is quite similar); the distribution is close to lognormal in all of the simulations with a median density of 
$\sim 100 \, {\rm cm^{-3}}$ and a broad dispersion of $\sim1.5\,$dex.   The highest resolved densities reach $>10^{6}\,{\rm cm^{-3}}$ in the ultra-high resolution simulation, but it is 
important to note that gas does not simply ``pile up'' gas at these high densities, which it does if we do not include feedback. We show this explicitly in Figure~\ref{fig:turb.highz.demo} by including 
the density PDF for a simulation with identical initial conditions, but no feedback \citep[see also the density distributions in the simulations without momentum-feedback in][]{teyssier:2010.clumpy.sb.in.mergers} -- in this case almost all the gas ends up at the maximum density allowed 
by our resolution ($\sim10^{6}\,{\rm cm^{-3}}$), with a small tail at low densities.  With feedback included, however, most of the mass is in GMC-like structures, but within those structures feedback ensures that most of the mass is in a more diffuse phase, rather than in the densest star-forming cores. 
The same conclusions pertain to our other galaxy models, but with lower median densities as expected; 
the volume-averaged $\langle n\rangle\sim1\,{\rm cm^{-3}}$ in the MW and Sbc models, with much of the mass in the star-forming disk in GMCs with a mean $\langle n \rangle \sim 10-30\,{\rm cm^{-3}}$ (and a resolved 
tail up to $\sim10^{6}\,{\rm cm^{-3}}$).
We caution that the distribution of low-density gas ($n\ll 1\,{\rm cm^{-3}}$) can be strongly altered by other sources of energetic feedback, such as SNe, stellar winds, and photo-ionization; the most dense gas, however, is where radiation pressure is likely to be most important.

The bottom panels in Figure~\ref{fig:turb.highz.demo} quantify the magnitude of the stellar feedback:   we show the integrated momentum supplied to the gas as a function of time, 
the typical initial velocity of the kicks as they are imparted to particles at the star-cluster scale, and the momentum-weighted optical depth of the gas clumps where the feedback is applied.

\begin{figure*}
    \centering
    \plotsidesmall{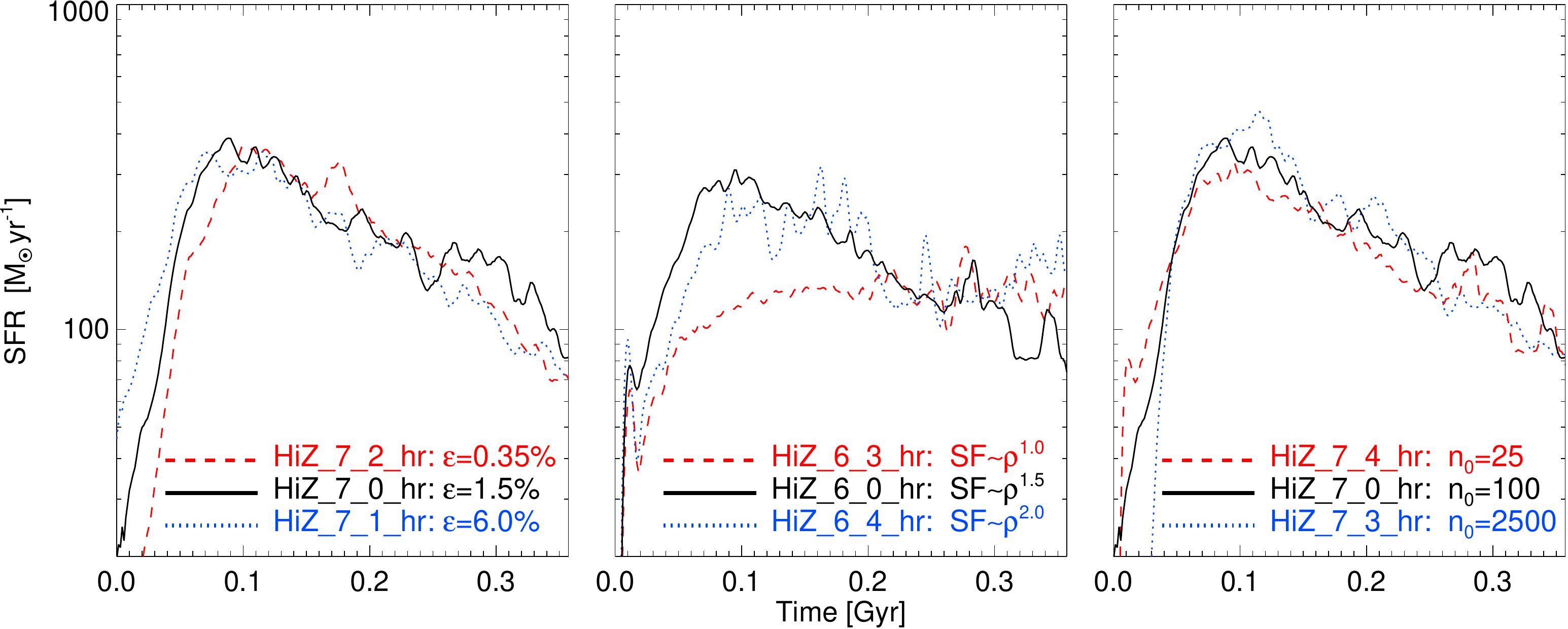}
    \caption{Star formation rate as a function of time for the HiZ model for variations in the {\em small-scale} (high-density) star formation law; all runs use our fiducial feedback parameters ($\eta_p = \eta_v = 1$).  These results demonstrate that the global star formation rate depends only weakly on the small-scale star formation law.    In each panel, the black solid line shows the standard star formation model:   $\dot{\rho}_{\ast}=\epsilon\,\rho/t_{\rm ff}$ above a threshold density $n_{0} = 100$ cm$^{-3}$, with $\epsilon=0.015$ and $t_{\rm ff}=\sqrt{3\pi/32\,G\,\rho}\propto \rho^{-0.5}$. 
    {\em Left:} Variations in the star formation efficiency $\epsilon$.
    {\em Middle:} Variations in the density-power law of the star formation model:   $\dot{\rho}_{\ast}\propto\rho^{n}$ with $n=1,\,1.5,\,2$, normalized so that $\dot{\rho}_\ast$ is the same as the default model at $n_{0}$.   {\em Right:} Variations in  the threshold density for star formation $n_0$.
     \label{fig:sfh.sflaw}}
\end{figure*}

\begin{figure}
    \centering
    \plotone{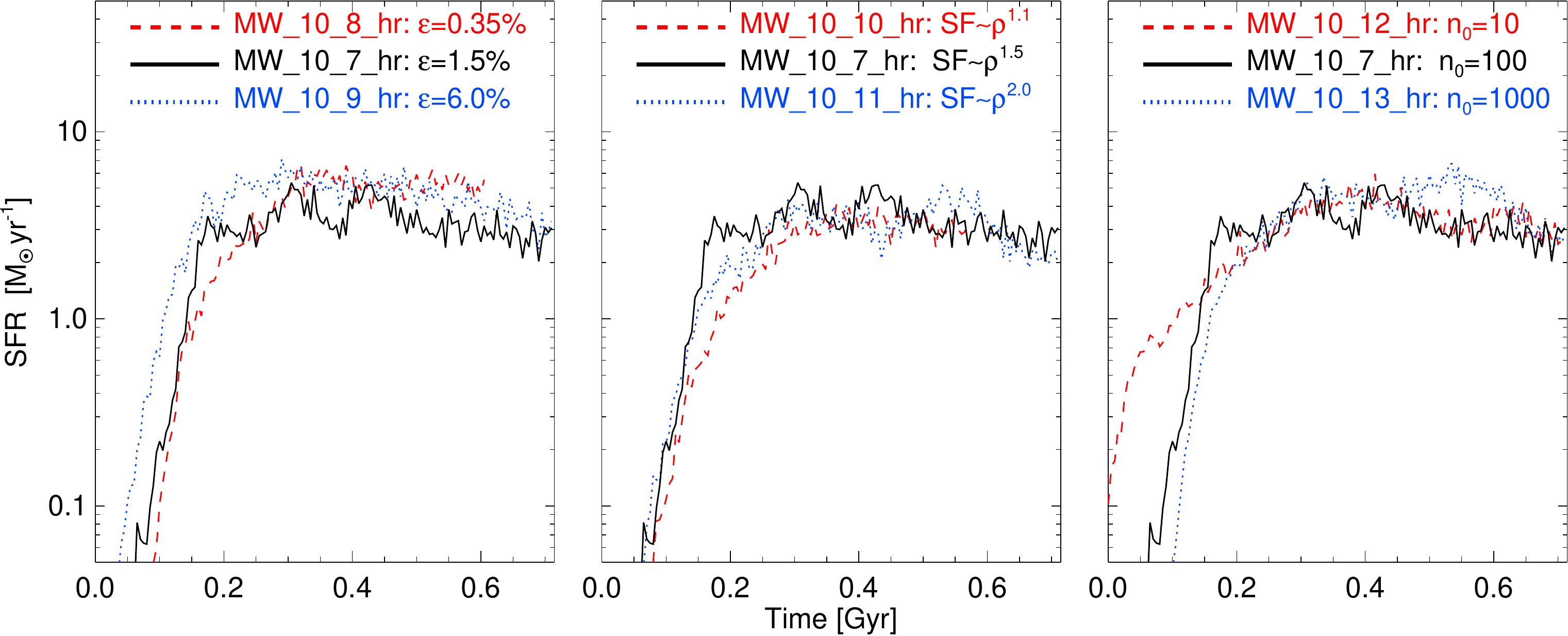}
    \caption{As Figure~\ref{fig:sfh.sflaw}, but for the MW model. 
    Again the global SFR is independent of the local, high-density SF law.
     \label{fig:sfh.sflaw.mw}}
\end{figure}

The values of the ``initial'' kick velocities given to the particles are large for the HiZ model, but not surprising given the very large star cluster masses ($\sim10^{8}\,\msun$) associated with the giant clumps in gas-rich high redshift systems \citep[see][]{murray:molcloud.disrupt.by.rad.pressure,genzel2011:clumps}; by contrast, the initial kicks in the MW-like system are much lower, tens of ${\rm km\,s^{-1}}$. Note also that the initial kick velocities are much larger than the velocity dispersion in the galaxy (in both cases): this occurs because the particles immediately interact with the surrounding ISM and share their momentum.   In \S \ref{sec:results:sfh:model} we show that for this reason, the choice of the initial kick velocity -- or even whether to continuously accelerate rather than ``kick'' particles -- is largely irrelevant.\footnote{For the same reason, the total fraction of gas particles initially ``kicked,'' which in these models is about $\sim10\%$, is unimportant.}   Instead, the important parameter is the total momentum supplied to the gas. 

With the large SFR and reasonably large kick velocities of the HiZ model, we might expect a sizeable super-wind to be generated by this feedback mechanism alone. However, in fact, the amount of mass in a proper super-wind (defined as e.g.\ the mass flux at $>V_{c}$ escaping to at least $\sim20\,$kpc) is relatively small relative to the SFR, about $\sim10\%$. 
This is because, as described above, the momentum is coupled in extremely dense regions and so rapidly shared among the gas particles. This can maintain a large velocity dispersion in the disk, but will not efficiently launch gas well out of the disk. Occasionally some material has an un-obstructed sightline out of the disk and escapes, but even then, the launch velocities are typically below the circular velocity, so the material is lofted up above the disk and then returns rapidly. It is likely that rather than winds being launched directly out of the galaxy from individual star clusters, some continuous acceleration mechanism is needed to act on gas once this local mechanism pushes it above the disk, in order to accelerate it out of the galaxy halo. This could be either pressure acceleration from hot, SNe-heated gas, or continuous radiation acceleration from the light which escapes the dense, optically thick regions we model here. In future work, we will investigate the properties of the galactic super-winds in more detail, and examine how these mechanisms interact with the feedback mechanism described here.

Figure~\ref{fig:turb.highz.demo} shows that the total momentum supplied to the gas is significantly larger than $\int\,(L/c)\,{\rm d}t = E_{\rm rad}/c$ because of the non-zero optical depths (where $E_{\rm rad}$ is the total radiated energy).  In fact, the numerical results are consistent with the total momentum supplied being given by $\simeq \eta_{p}\,\langle \tau_{\rm IR} \rangle E_{\rm rad,\,young}/c$, where $E_{\rm rad,\,young}$ is the integrated luminosity from young stars with ages $<10^{6}$\,yr.  This is because the feedback begins to disperse the densest regions on a $\sim 10^6$ yr timescale.   Figure~\ref{fig:turb.highz.demo} also shows the median and dispersion in the clump optical depths for the regions where the feedback is applied:   $\tau_{\rm IR} \sim 50$ in the highest resolution simulation.   This corresponds to gas surface densities $\sim 10$ g cm$^{-2}$, similar to the observed surface densities of massive star clusters.   The average $\tau$ over the entire disk is, of course, significantly smaller, $\tau_{\rm IR} \sim 0.1-1$. For this reason, for the MW, Sbc, and SMC models, although the disk-averaged $\tau_{\rm IR}$ is significantly smaller than the HiZ model, their ``effective'' $\tau_{\rm IR}$ is not too much smaller. Despite the global gas mass being lower, the mass that actually forms stars and star clusters tends to be compact cores at high three-dimensional ($n\gtrsim10^{4}\,{\rm cm^{-3}}$) and surface densities ($\Sigma\gg 1000\,{\msun\,{\rm pc^{-2}}}$).

Figure~\ref{fig:turb.highz.demo} shows that the total momentum input does not depend that strongly on resolution or on $\eta_p$ at a given resolution.   In fact, the optical depths decrease with increasing $\eta_p$, maintaining approximately the same total momentum input.   Physically, this is because for more/less efficient feedback the gas collapses to lower/higher densities (on average).  The fact that the total momentum supplied by feedback depends only weakly on resolution and $\eta_p$ is a consequence of the disk self-regulating to maintain $Q \sim 1$.   This constraint picks out a $\delta v$ as a function of $\Sigma_g$ and $\Omega$ (eq. \ref{eqn:Q}) -- the momentum input then adjusts to produce the required $\delta v$. Again, the same is true in all galaxy models. 

The column density distribution within individual star forming clumps can strongly influence the efficacy of radiation pressure feedback.   For example, if there is a very broad distribution with a large number of optically thin sightlines,  then even though the average column density of a clump in the IR may be large, a sizeable fraction of photons would leak out of optically thin sightlines and the effective optical depth for the purposes of feedback would be reduced ($\eta_{p}<1$, in our parameterization). 
To quantify this, we considered a number of massive clumps in our HiZ simulation and determined the optical depth along $\sim1000$ sightlines 
evenly spaced in solid angle from the clump center outwards (following the methodology in 
\citealt{hopkins:lifetimes.letter}).  The characteristic dispersion in optical depth {\em for a given} clump is small, $\sim0.2$\,dex, similar to what has been found in smaller-scale simulations of 
individual clouds \citep{ostriker:2001.gmc.column.dist} 
and measured observationally \citep{wong:2008.gmc.column.dist,goodman:2009.gmc.column.dist}.\footnote{In detail, we find there is a narrow core in the distribution and a broader-than-lognormal wing; the distribution is better fit by an exponential at high columns, with   $P(\log{N_{H}})\propto \exp{[-|\log{N_{H}/N_{0}}|/0.22]}$.   This is broadly consistent with randomly distributed ``patchy'' obscuration within clouds.}  Note that this dispersion for an individual star-forming clump is much smaller than for the galaxy as a whole.      A key consequence of the relatively narrow column density distribution within clumps is that only a negligible fraction of the sightlines are optically thin enough for the photons to rapidly leak out; it thus appears reasonable to use the mean clump column density when quantifying the feedback produced by the IR radiation (as we assume in our fiducial models; see \S \ref{sec:stellar.model:momentum}).

It is also straightforward to calculate the total momentum deposition or infer it from the momentum coupled 
and typical velocities, $\dot{E}_{w} \approx (1/2)\,\dot{p}_{w}\,v_{w}$. For the values in Fig.~\ref{fig:turb.highz.demo}, this is  about $0.2-0.4\%$ of the stellar luminosity. Interestingly, recent observations of massive stars \citep{freyer:2006.massive.star.wind.egy} and star forming regions \citep{lopez:2010.stellar.fb.30.dor} have suggested similar $\sim1\%$ efficiencies for transfer of luminous energy into bulk motions and (post-shock) thermal energy. 

In future work (in preparation), we will compare the structural properties of the ISM and dense gas in these simulations and local observed galaxies in detail. However, in low-density gas, typical of much of the mass in the SMC model and the intermediate/diffuse phases in the Sbc and MW models, it is likely that other processes (shock-heating by SNe and stellar winds, photo-ionization, magnetic fields) can play a significant role in shaping the gas dispersions and density distribution. We therefore defer a more detailed comparison with these observations until the models include some of these effects. However, preliminary experiments show that while the systematic values discussed above can shift, the qualitative conclusions remain intact as other feedback mechanisms are introduced.

\vspace{-0.4cm}
\section{Dependence on Model Parameters}
\label{sec:results:sfh}

In this section we show that the results summarized in \S \ref{sec:results:global} do not depend sensitively on the assumed local star formation law (\S \ref{sec:results:sfh:sflaw}) or the precise feedback parameters adopted (\S \ref{sec:results:sfh:model}).  We focus on the star formation history when presenting these results.  We again focus on the HiZ model since it is the most self-gravitating and therefore its SFH tends to be the most sensitive to variations in the simulation parameters.  However, we carried the same experiments for the MW-like simulation and found comparable results, which are also shown below. In the appendix we show that our results also do not depend strongly on how we numerically implement the stellar feedback.

\begin{figure*}
    \centering
    \plotside{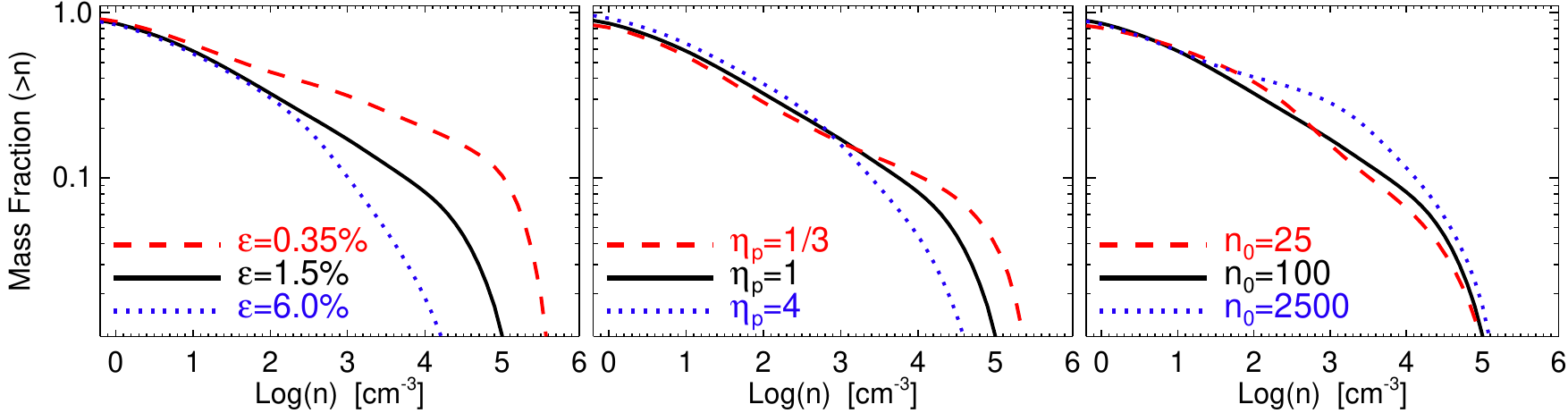}
    \caption{Cumulative gas mass fraction above a given density $n$ 
    for the (high-resolution) HiZ model for variations in the small-scale star formation law (Fig.~\ref{fig:sfh.sflaw}) 
    and feedback efficiency (Fig.~\ref{fig:sfh.fb}). 
    {\em Left:}  Density distribution for different values of the star formation efficiency $\epsilon$: for smaller (larger) $\epsilon$, more (less) mass must collapse to high densities for the star formation to self-regulate 
    (the high-$\rho$ cutoff is set by resolution limits).
      {\em Middle:} Density distribution for different values of the momentum deposition per unit star formation ($\eta_{p}$; eq.~\ref{eqn:mom.vs.L}). For larger $\eta_p$, gas is more efficiently removed from dense regions.   {\em Right:} Density distribution for different values of the threshold density for star formation $n_0$. Larger $n_{0}$ requires that the gas collapse to somewhat higher densities before the star formation can self-regulate.     
    \label{fig:density.vs.fb}}
\end{figure*}

\subsection{Dependence on the Local SF Law}
\label{sec:results:sfh:sflaw}

Figures~\ref{fig:sfh.sflaw}-\ref{fig:sfh.sflaw.mw} shows how the star formation history in feedback-regulated simulations  depends on the {\em local} star formation prescription used at high densities. For our fiducial $\eta_p = \eta_v = 1$ model, Figures~\ref{fig:sfh.sflaw}-\ref{fig:sfh.sflaw.mw}  vary the star formation efficiency in dense gas $\epsilon$, the power-law slope of the star formation law, and the threshold density for star formation $n_0$ (eq. \ref{eqn:sfr}).

The key result in Figures~\ref{fig:sfh.sflaw}-\ref{fig:sfh.sflaw.mw} is that there is very little dependence of the SFH on the high density star formation law.   Specifically, Figure~\ref{fig:sfh.sflaw} ({\em left panel}) shows results for 
our canonical value of $\epsilon = 1.5\%$, a larger value of $6\%$, and a smaller value of  
$\epsilon = 0.35\%$ (we have also examined several intermediate values).  
This range of $\epsilon$ corresponds to a factor of $20$ different 
star formation timescale at fixed density.   We find, however, at most $\sim 30 \%$ differences in the SFR  once the system has reached approximate equilibrium. In the MW-like model (Figure~\ref{fig:sfh.sflaw.mw}) the conclusion is identical.
Secondly, we vary the power-law index of the local SF law ({\em middle panel}). In our canonical implementation, $\dot{\rho}_{\ast}\propto \rho/t_{\rm dyn}\propto \rho^{1.5}$; we compare this to simulations with $\dot{\rho}_{\ast}\propto \rho/t_{0}\propto \rho^{1.0}$ and $\dot{\rho}_{\ast}\propto \rho^{2.0}$, normalized such that $\dot{\rho}_{\ast}$ is identical at the threshold density $\rho_{\rm 0}$.   There are early-time differences in the star formation histories, but given the magnitude of the change to the star formation prescription the results are broadly similar. The biggest change 
appears when $\dot{\rho}_{\ast}\propto \rho$, i.e.\ when the gas consumption timescale is constant, independent of density; in this regime, the gas cannot necessarily be consumed quickly on small 
scales, so the collapse from large to small scales is no longer the dominant rate-limiting step in star formation (a slightly larger exponent, e.g.\ $\dot{\rho}_{\ast}\propto \rho^{1.2}$, much more closely resembles the canonical $\propto \rho^{1.5}$ case). We show this for the MW-like model in Figure~\ref{fig:sfh.sflaw.mw}, comparing 
$\dot{\rho}_{\ast}\propto \rho^{1.5}$, $\dot{\rho}_{\ast}\propto \rho^{2.0}$, $\dot{\rho}_{\ast}\propto \rho^{1.1}$. 
The relative differences in the $\dot{\rho}_{\ast}\propto \rho^{2.0}$ are even smaller, and making the 
exponent just slightly super-linear ($\propto \rho^{1.1}$) gives a nearly identical SFR to $\propto \rho^{1.5}$.
Finally, we vary the SF density threshold $n_{0}$ ({\em right panel}); from our canonical value of $100\,{\rm cm^{-3}}$, we also consider  $n_0 = 25\,{\rm cm^{-3}}$ and $n_0 = 2500\,{\rm cm^{-3}}$ (with other intermediate values sampled as well).   At early times, before the initial conditions have been replaced by a self-consistent equilibrium, the SFR is higher with a lower threshold (unsurprisingly).   However, once enough time has elapsed for gas to collapse to high densities and initiate significant feedback, the SFHs are again nearly identical despite a factor of $16$ change in the threshold for star formation. The same result obtains in the MW model, for which (given the lower mean density) we vary $n_{0}=10-1000$. 
Moreover, for a MW-like model with similar resolution, \citet{saitoh:2008.highres.disks.high.sf.thold} find the same 
result in a more limited study varying the small-scale star formation efficiency, 
but with a different simulation algorithm and different feedback mechanism (SNe) implemented.
We have also repeated these experiments for different values of $\eta_{p}$ ($=1/3,\,4,\,10$) 
and $\eta_{v}$ ($=2$ and continuous acceleration), and reach similar conclusions in each case.

Our interpretation is that the weak dependence of the global star formation rate on the small scale star formation model is a consequence of the turbulence driven by stellar feedback, and the self-regulation to $Q \sim 1$ (see, e.g., \citealt{thompson:rad.pressure}). Specifically, gravity causes gas to collapse to high density, where some of it forms stars, while most of the gas is driven back out to lower densities by feedback.  The key step that regulates the star formation rate is this cycle of collapse and expulsion, which has a timescale $\sim$ the global dynamical time of the galaxy -- this is also the decay timescale for large-scale turbulence in the galaxy. The details of feedback on small scales should also not be important, so long as it is sufficient to self-regulate \citep[compare 
our result to][]{saitoh:2008.highres.disks.high.sf.thold}.
So long as the star formation timescale at the threshold density is small compared to the global dynamical time (i.e.,  $\epsilon$ not too small and $\rho_0 > $ the $\langle \rho \rangle$ of the galaxy) and the threshold is well-resolved numerically (i.e., $\rho_0$ is not too large), the star formation rate is insensitive to the details of the small-scale star formation law. 

More generally, if the support needed to maintain stability against runaway star formation is set by the luminosity/mass in young stars, the SFR can self-regulate to $Q \sim 1$.  For example, if the SFR set by the small-scale physics is too low to maintain $Q \sim 1$ given the large-scale conditions, gas simply collapses further to slightly higher densities until the required feedback power is generated, sufficient to halt further collapse.   The high density star formation law thus determines some of the properties of the high density gas, but not the global SFR.

Figure~\ref{fig:density.vs.fb} supports this interpretation by showing the cumulative gas density distribution (mass fraction $>n$) for different values of $\epsilon$ ({\em left panel}), $n_0$ ({\em right panel}), and the feedback parameter $\eta_p$ discussed in the next section ({\em middle panel}).
Figure~\ref{fig:density.vs.fb} shows that when the high-density star formation efficiency $\epsilon$ is smaller (larger), the gas distribution adjusts so that there is more (less) mass at high densities, so as to produce a similar total SFR (as in Fig.~\ref{fig:sfh.sflaw}.)  When the threshold density $n_{0}$ is varied, the mass at high densities shifts accordingly.   For example, increasing $n_0$ causes the gas that would have formed stars at the previous threshold to collapse to somewhat higher densities before it begins to form stars. 

Note that the mass at low densities is nearly unchanged -- the disks are not in global collapse (they are regulated by feedback), but the gas locally collapses to the densities needed to maintain the same SFR. For this reason, the Schmidt law predicted by each of the models in Figure~\ref{fig:sfh.sflaw} is nearly identical. They have the same range in surface densities (set by the initial conditions and exhaustion via star formation, which must be the same since they have the same star formation history), and so self-regulate at the same SFR.

\citet{schaye:2010.cosmo.sfh.sims}, using much lower resolution cosmological simulations, also find a galaxy wide star formation rate that is independent of the details of the small scale star formation law employed. 
However, in their case, because star formation laws are applied globally (on $>$kpc scales), it is the global 
gas mass that self-adjusts (e.g.\ lowering the star formation efficiency leads to inflows larger than the SFR building up the global gas mass until the SFR is similar to the cosmological inflow rate), so the systems do not necessarily obey the observed Schmidt-Kennicutt relation. In our case, neither the SFR nor global gas mass varies; what {\em does} alter is the gas fraction at the very highest densities available to the simulations.

\begin{figure*}
    \centering
    \plotsidesmall{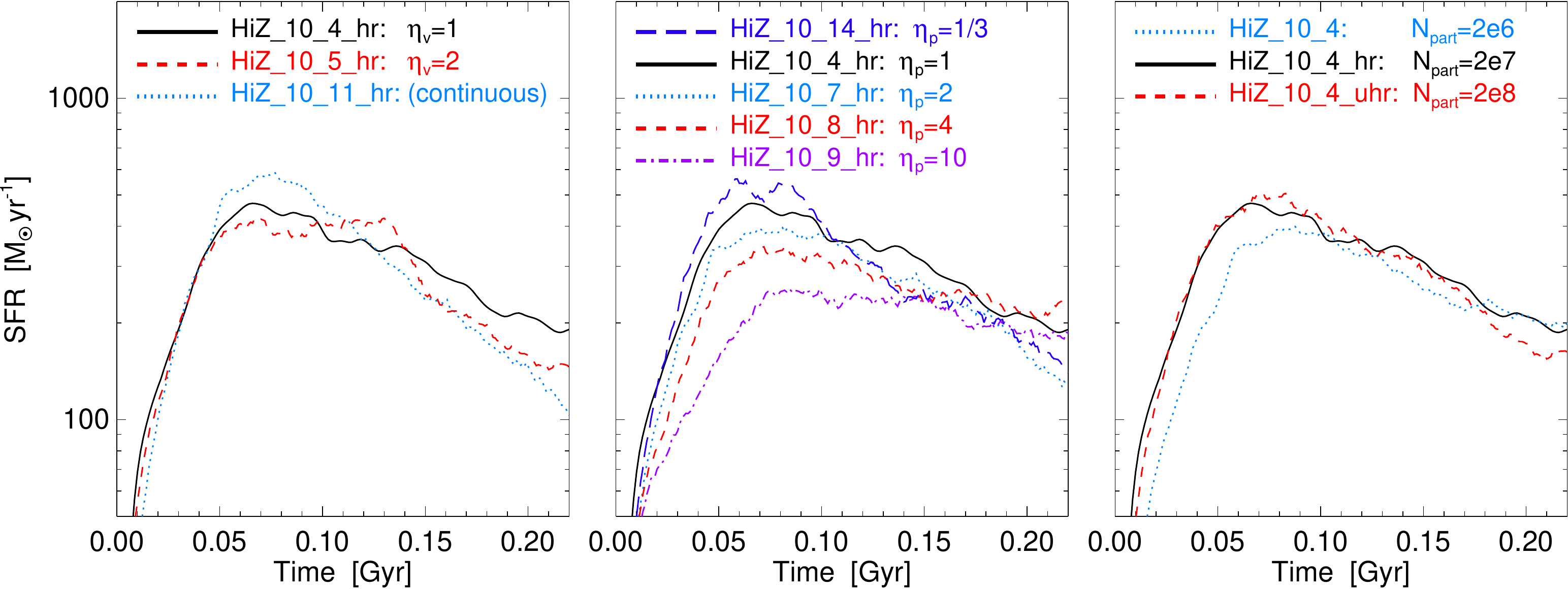}
    \caption{Star formation rate as a function of time for the HiZ model for different feedback parameters and resolution.    {\em Left:} Variations in the initial kick velocities at fixed 
    momentum-loading.  We also show a simulation with stochastic particle "kicks" replaced by continuous acceleration of all particles.   In all of these simulations, the gas shares its momentum with the rest of the surrounding clump and thus produces similar dynamics.
    {\em Center:} Variations in the momentum deposition per unit star formation ($\eta_{p}$; eq.~\ref{eqn:mom.vs.L}).    The star formation rate decreases by less than a factor of 2 over a factor of 10 in $\eta_p$.  {\em Right:} Variations in resolution.
    Increasing the particle number from $\sim10^{6}$ to $\sim10^{7}$
    (our ``intermediate'' vs ``high'' resolution) increases the star formation rate at early times 
    by a moderate amount ($\sim20-40\%$). But after about one orbital time, the results are quite similar.  
    Going to yet higher resolution gives nearly identical results. 
    \label{fig:sfh.fb}}
\end{figure*}

\begin{figure}
    \centering
    \plotone{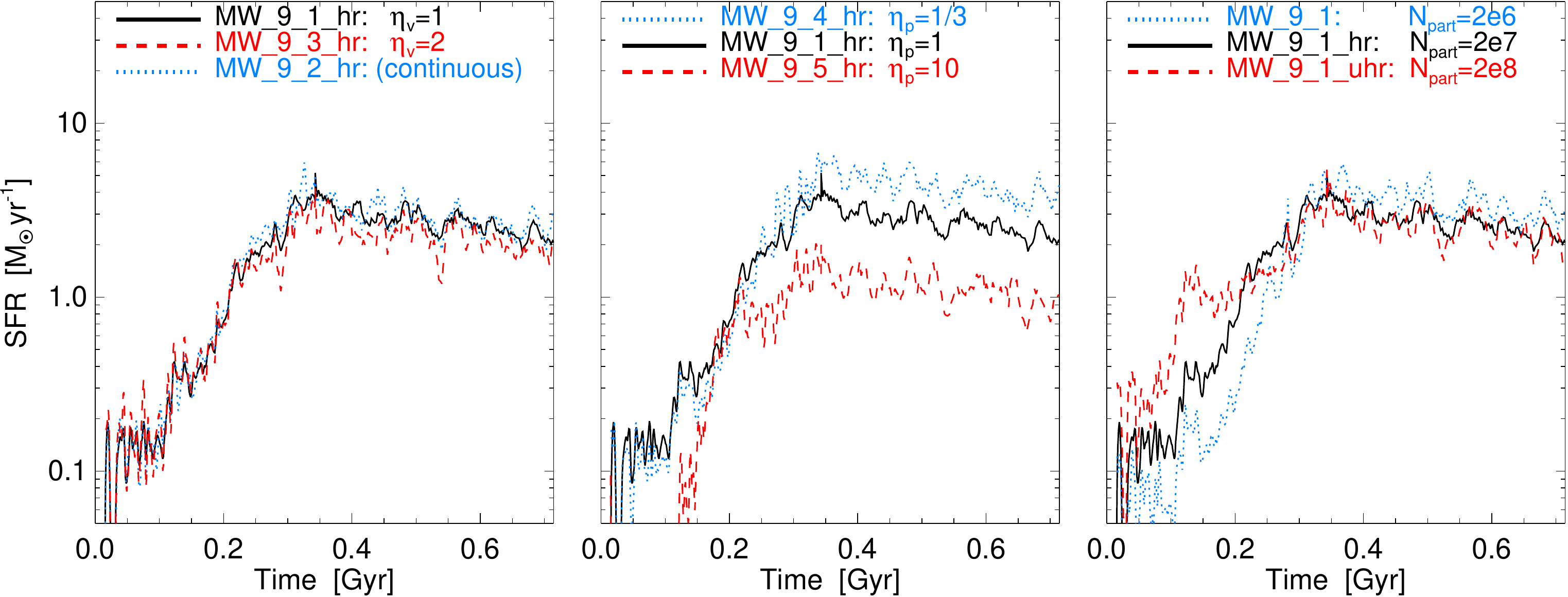}
    \caption{As Figure~\ref{fig:sfh.fb}, but for our MW model. 
    Dependence on $\eta_{v}$ and resolution 
    is even smaller than the HiZ case; the dependence on $\eta_{p}$ is similar.
    \label{fig:sfh.fb.mw}}
\end{figure}

\vspace{-0.3cm}

\subsection{Dependence on the Feedback Efficiency}
\label{sec:results:sfh:model}

Figure~\ref{fig:sfh.fb} (Figure~\ref{fig:sfh.fb.mw}) shows how the star formation history of our HiZ (MW) model depends on the feedback parameters $\eta_p$ and $\eta_v$ (eqs. \ref{eqn:mom.vs.L} \& \ref{eq:v}) and on whether we implement the momentum-feedback continuously or via kicks ({\em left and middle panels}).    All of the variations are with respect to our standard $\eta_p = \eta_v = 1$ model.

Figure~\ref{fig:sfh.fb} ({\em left panel}) shows that simulations with $\eta_v = 1$ and $\eta_v = 2$ produce very similar SFHs.   These two simulations both have $\eta_p = 1$ and thus have the same momentum injection rate.   Physically, the similarity in their SFHs arises because the particles interact with ambient gas and share their momentum efficiently.   The end result is that clumps being destroyed by stellar feedback have velocities comparable to the escape velocity from the clump, relatively independent of the initial velocities we input.   Figure~\ref{fig:sfh.fb} also shows a comparison of two different methods of implementing the same momentum flux:   particle `kicks' and continuous acceleration (see \S \ref{sec:stellar.model:momentum}).   It is reassuring that these two methods produce quite similar results -- this again highlights that the critical parameter is the rate at which momentum is deposited into the ISM, not precisely how it is deposited. The MW-like model in Figure~\ref{fig:sfh.fb.mw} gives identical conclusions (the dependence is even weaker in this case).

Figure~\ref{fig:sfh.fb} ({\em middle panel}) also compares simulations with varied momentum-injection per unit star formation $\eta_{p}$, from $\eta_p = 0.3-10$ (all at fixed $\eta_v = 1$).   As expected, the quasi-steady SFR decreases as the efficiency of momentum-injection increases.   However, the decrease in the SFR is rather mild, with a factor of $\sim 2$ change in SFR over a factor of $10$ in $\eta_p$. 
The same scaling holds in the MW-like model in Figure~\ref{fig:sfh.fb.mw}.
  Naively one might expect an inverse linear scaling $\dot{M_*}\propto \eta_{p}^{-1}$ \citep{thompson:rad.pressure}.   Specifically, the turbulent energy dissipation rate per unit area is
\begin{equation}
\label{eqn:dPdA}
\frac{{dE}}{{dA}\,{dt}}\sim \langle \Sigma \rangle \,\delta v^2\,\Omega \sim G\,\langle \Sigma \rangle^{2} \delta v \end{equation}
where $\langle \Sigma \rangle$ is the mean surface density of the disk and we have assumed $Q \sim 1$ and that turbulence dissipates on a crossing time $\sim h/\delta v \sim \Omega^{-1}$.   The total momentum injection rate scales with the 
SFR as $\dot{P} \sim (1 + \eta_{p}\,\tau_{\rm IR})\,L/c 
\sim (1 + \eta_{p}\, \Sigma \,\kappa_{\rm IR})\,\epsilon_{\ast}\,\dot{M}_{\ast}\,c$,
where $\epsilon_{\ast}\approx 4\times10^{-4} = L/\dot{M}_{\ast}\,c^{2}$.  If the momentum-injection ultimately drives turbulent motions with random velocity $\delta v$, the associated energy injection rate is $\sim \dot P \delta v$.   Balancing this against the dissipation in equation~\ref{eqn:dPdA}, we find that the star formation rate per unit area $\Sigma_{\rm SFR}$ is given by
\begin{equation}
\label{eqn:selfreg.expected}
\Sigma_{\rm SFR} \sim \frac{G\,\langle \Sigma \rangle^{2}}{\epsilon_{\ast}\,c \, (1+\eta_p \, \tau_{\rm IR})}.
\end{equation}
For parameters relevant to Figure~\ref{fig:sfh.fb}, $\eta_p \tau_{\rm IR} > 1$ and so equation \ref{eqn:selfreg.expected} implies that $\dot M_* \propto (\tau_{\rm IR} \eta_p)^{-1}$.    This does not, however, imply $\dot M_* \propto \eta_p^{-1}$.  One reason is that equation \ref{eqn:selfreg.expected}  neglects the possibility of significant cancellation in colliding/canceling flows.   More importantly, however, for (say) larger $\eta_p$, the fraction of mass at high densities decreases because feedback is more effective. This is shown explicitly in Figure~\ref{fig:density.vs.fb} ({\em middle panel}): as $\eta_{p}$ increases, the density distribution cuts off more sharply at high $n$. As a result, the optical depth $\tau_{\rm IR}$ in the regions of massive star formation decreases.  This demonstrates that the momentum input $\propto \eta_p \tau_{\rm IR}$, and thus the star formation rate, must scale sub-linearly with $\eta_p$ (as in Fig. \ref{fig:sfh.fb}).    Assuming that feedback removes gas from high to low density at a rate $\propto \eta_p$, we would expect the fraction of mass at high densities -- and the optical depth in those regions --  to decrease roughly as $\eta_p^{-1}$.   This is why both the momentum input $\propto \eta_p \tau_{\rm IR}$ and the star formation rate $\propto (\eta_p \tau_{\rm IR})^{-1}$ have only a weak dependence on $\eta_p$.   This property of our numerical simulations is one of the most significant differences between our results and previous analytic treatments of star formation regulated by radiation pressure \citep{thompson:rad.pressure}.   It is important to stress that this self-regulation to achieve the same star formation rate relatively independent of the feedback parameter $\eta_p$ is only a property of models in which the momentum injection rate is proportional to the gas surface density (eq. \ref{eqn:mom.vs.L}); that is, it is only a property of feedback by radiation pressure, not momentum injection associated with supernovae or stellar winds.   

\vspace{-0.3cm}
\subsection{Dependence on Resolution}  

Figures~\ref{fig:sfh.fb}-\ref{fig:sfh.fb.mw} ({\em right panel}) shows how the star formation histories of our fiducial $\eta_p = \eta_v = 1$ HiZ and MW models depend on particle number, with $N_{\rm part} = 2 \times 10^6, \, 2 \times 10^7$, and $2 \times 10^8$.   The basic evolution of the SFR is very similar in all cases.   The SFR is $\sim 25-40 \%$ higher at early times in the $N_{\rm part} = 2 \times 10^7$ simulation relative to $N_{\rm part} = 2 \times 10^6$, but there is a much smaller change going to yet higher resolution.   Moreover, after a few dynamical times, all of the simulations have a comparable SFR.   We find similar convergence for different galaxy models and different feedback parameters.

\vspace{-0.4cm}
\section{The Global Schmidt-Kennicutt Law}
\label{sec:results:ks}

\begin{figure*}
    \centering
    \plotsidesmallestest{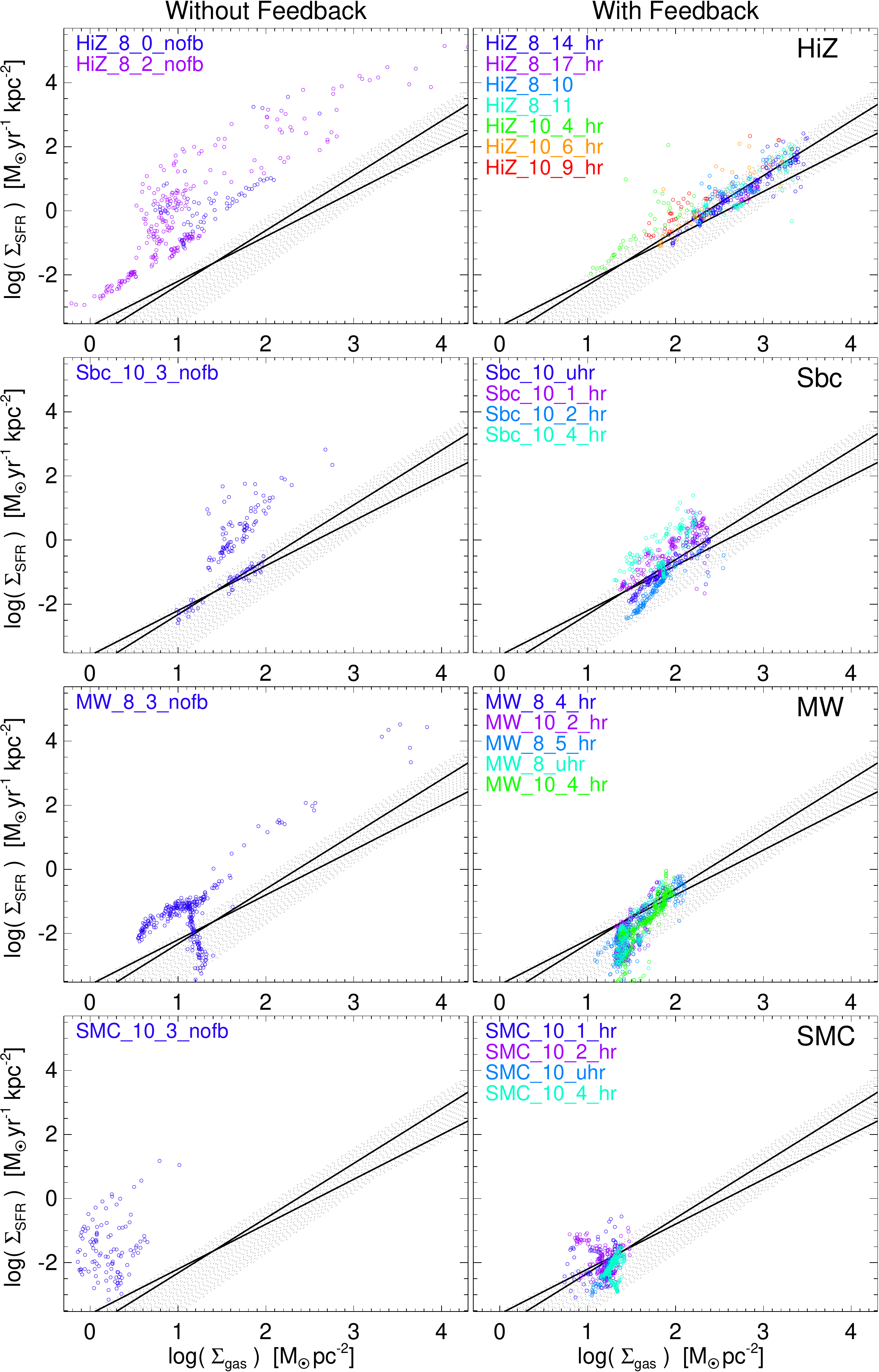}
    \caption{The global Kennicutt-Schmidt relation between star formation rate density and gas surface density in our simulations. {\em Left:} Without feedback. {\em Right:}  Simulations with our feedback model for a range of parameters (see Table~\ref{tbl:sims}).  Each row is a different galaxy model (see Table~\ref{tbl:sim.ics}). In each panel, a point corresponds to one time snapshot in the simulation, evenly spaced in $\sim10^{6}\,$yr intervals (starting after two dynamical times). The surface densities are as viewed face-on, averaged within the circular radius that encloses $1/2$ of the star formation. Solid lines show the fits to the data in \citet{kennicutt98} and updated with high-redshift galaxies by \citet{genzel:2010.ks.law}. Grey shaded region shows the $90\%$ completeness range at each gas surface density from the compilation of the systems observed in those two works as well as the compilations in \citet{bigiel:2008.mol.kennicutt.on.sub.kpc.scales} and \citet{daddi:2010.ks.law.highz}. Without feedback, the gas experiences runaway collapse  and is consumed in less than a dynamical time, predicting star formation rate densities in excess of the observed KS-relation by factors of $\sim100-10^{4}$.  With feedback, the gas disks quickly self-regulate and reach an approximate equilibrium comparable to that observed.\label{fig:ks}}
\end{figure*}

Figure~\ref{fig:ks} compares the global Kennicutt-Schmidt law 
predicted by simulations with and without our stellar feedback 
model. We measure $\Sigma_{\rm SFR} \equiv \dot{M}_{\ast}(<R_{\rm sfr})/\pi\,R_{\rm sfr}^{2}$ 
as a function of $\Sigma_{\rm gas} \equiv M_{\rm gas}(<R_{\rm sfr})/\pi\,R_{\rm sfr}^{2}$, 
where $R_{\rm sfr}$ at each time is defined as the half-SFR radius  via $\dot{M}_{\ast}(<R_{\rm sfr})=\dot{M}_{\ast}/2$. This radius is chosen to loosely correspond to the 
half-optical or half-H$\alpha$ radii used in various observational studies, 
but adopting a different choice (e.g.\ the half-gas mass radius) primarily shifts the models along the relation. The numerical results are shown every Myr.  The numerical results in Figure~\ref{fig:ks} are compared with several different observational inferences of the K-S relation:   the best-fit power-law relations from 
low-redshift in \citet{kennicutt98} and high-redshift in \citet{bouche:z2.kennicutt,
genzel:2010.ks.law}, 
together with the $10-90\%$ interval of all points from the combined compilations in those studies as well as 
\citet{bigiel:2008.mol.kennicutt.on.sub.kpc.scales} 
and \citet{daddi:2010.ks.law.highz} (the shaded range).\footnote{Note 
that the shaded range falls below the best-fit power-laws at low 
surface densities because the power-law fits did not study the low surface-density `cutoff' due to a low molecular fraction.} The models with and without feedback have identical initial conditions in each case.   

Without feedback our simulations predict a SFR surface density well in excess of the K-S law (see also Fig.~\ref{fig:sfh.ov}).   Absent feedback, the gas cools and cannot avoid runaway collapse; most of the gas is consumed into stars in a single dynamical time, leading to SFR surface densities 
$\sim 10-10^{4}$ times larger than observed. 

In contrast, the simulations with feedback lie close to the observed relation at essentially 
all times.   This is true over the range of feedback parameters and resolution we have studied; 
the runs in Figure~\ref{fig:ks} span a range in $\eta_{p}=0.5-10$, 
$\eta_{v}=1-4$, star formation law variations as in Figure~\ref{fig:sfh.sflaw}, and resolution 
($N_{\rm particles}\sim10^{5}-10^{8}$). 
Varying the simulation parameters for each galaxy model tends to shift  the systems along the KS relation, rather than dramatically off the relation.   For each galaxy model, there is a broad dynamic range in $\Sigma$ 
covered; in particular, the high-z simulations lie on the observed 
relation over multiple decades in density. 
The average slope of the relation (if we consider all galaxies together) is quite similar to 
that observed; however we see that there can be significant variation in the slope within galaxies, 
also commonly observed \cite[see][]{bigiel:2008.mol.kennicutt.on.sub.kpc.scales}.\footnote{For example, 
the MW-like simulation is significantly steeper at low densities than the median relation, but quite 
similar to spatially resolved observations in M51 \citep{kennicutt:m51.resolved.sfr}. 
However, this is the regime where we expect other physics (e.g.\ the addition of SNe feedback 
and possibly the effects of detailed molecular chemistry) may become important.}
Altogether, the feedback-regulated simulations lie on the observed KS relation over a 
dynamic range from $\sim \Sigma_{\rm gas}\sim 10^{7}-10^{10}\,\msun\,{\rm kpc^{-2}}$. 
The scatter about the Schmidt law predicted is also similar to that observed, about $0.5$\,dex. 

\vspace{-0.3cm}

\section{Discussion}
\label{sec:discussion}

We have presented a new numerical method for treating stellar feedback in hydrodynamic simulations of galaxies.  We have implemented this method in the SPH code Gadget-3 but our approach is general and can be utilized in both Lagrangian and Eulerian codes (see \S \ref{sec:sims}).  Our stellar feedback model is motivated by the physics of feedback in dense environments:   under these conditions, gas cools rapidly and the primary dynamical influence of stellar winds, supernovae, and the stellar radiation field is the momentum they impart to the ISM.  In addition to formulating the general method, we have carried out a detailed study of the properties of this stellar feedback model in isolated (non-cosmological) disk galaxy simulations, from models motivated by massive $z \sim 2$ galaxies forming stars at $\sim 100-300 \, \msunyr$ to models of SMC-like dwarf galaxies.   These disk galaxy calculations are not intended to be quantitatively applicable to real systems; rather, they illustrate our method and demonstrate the critical importance of including stellar feedback by momentum injection.  In a future paper, we will combine the method developed in this paper with more standard treatments of supernova and stellar wind heating, to produce a more comprehensive stellar feedback model.

High-resolution numerical simulations of isolated galaxies and galaxy mergers, as well as cosmological ``zoom-in'' simulations, can readily resolve the formation of numerous dense gaseous clumps via gravitational instability (provided the cooling to low temperatures $\sim 100$ K is not artificially suppressed); we dub these clumps GMCs although we do not include the physics of molecule formation in our simulations.   In observations of nearby galaxies, most of the star formation occurs in GMCs -- this is also true in our simulations -- and thus it is important to have at least an approximate model for GMC disruption by stellar feedback.

To model stellar feedback, we implement an on-the-fly clump finding algorithm to identify high density star-forming clumps (i.e., GMCs).  We then deposit momentum into the surrounding gas at a rate proportional to the radiation produced by young stars in the clump; this force is directed radially away from the center of the GMC.  More precisely, the force we apply scales as $\sim \tau_{\rm IR} L/c$ (eq.~\ref{eqn:mom.vs.L}) where $\tau_{\rm IR}$ is the optical depth of the clump to IR photons and the stellar luminosity $L$ is calculated as a function of time given the stellar ages using Starburst99 models.    Although our model is quantitatively motivated by radiation pressure on dust,  the momentum flux from supernovae and massive stellar winds can also be significant.   We will study the relative importance of these different feedback mechanisms in detail in a future paper.

The model we have developed is distinct from the stellar feedback models used in most of the galaxy formation literature.    First, we input momentum, rather than thermal energy, into the ISM around young stars.   The motivation for this choice is that momentum, not energy, is the relevant conserved quantity in dense, rapidly-cooling gas.   Moreover, the feedback we implement scales with the local surface density of the GMCs, as expected for the radiation pressure produced by stellar photons as they are degraded by dust from the UV to the far IR (eq.~\ref{eqn:mom.vs.L}).  As summarized below and in \S \ref{sec:results:global}, we find that this surface density dependence is critical to the evolution of our galaxy models.

In our study of stellar feedback in isolated galaxies, we {\em do not} ``turn off'' hydrodynamic forces, cooling, star formation, and/or other physics in the gas to which the feedback is applied.  By contrast,  many stellar feedback implementations in the galaxy formation literature either turn off hydrodynamic forces in winds for some free-streaming length (typically such that winds escape the galaxy)  or turn off cooling and star formation in supernova-heated gas for some period of time.   In such models, the induced velocities on galactic scales are essentially determined by hand (through adjusting the relevant parameters) as is the presence/absence of a global galactic wind driven by stellar feedback.  In our model, the single key parameter is the momentum supplied to high density gas around star clusters -- the resulting {galaxy-wide} turbulence, the properties of galactic winds, etc. are all {\em predictions} of our model.   The model remains ``sub-grid,''  but on the scale of individual molecular clouds rather than the galaxy as a whole.     

We are able to directly model the stellar feedback without artificially modifying the underlying equations for several reasons.   The high resolution in our simulations allows us to partially resolve the multi-phase ISM structure: since star formation is spatially inhomogeneous, the stellar feedback is as well, which self-consistently maintains a turbulent and multi-phase ISM structure (Fig. \ref{fig:pics}).   Perhaps more importantly, the feedback is momentum-driven and the forces are directed away from the centers of local gas overdensities (GMCs), the sites of massive star formation. As a result, the feedback is effective even in dense regions of the ISM, in which the cooling time is much shorter than the dynamical time.   In standard treatments of feedback by supernovae, the feedback is inefficient in dense regions because the thermal energy supplied by supernovae is rapidly radiated away
\citep{thackercouchman00,governato:disk.formation,
ceverino:cosmo.sne.fb,bournaud:2010.grav.turbulence.lmc,
teyssier:2010.clumpy.sb.in.mergers,brook:2010.low.ang.mom.outflows}. 
Thus many simulations that nominally include stellar feedback do not in fact have feedback that is quantitatively of the correct order of magnitude.

As a first assessment of the implications of our new stellar feedback model, we have used it to study star formation in a wide range of (non-cosmological) disk galaxy models, representing systems ranging from SMC-like dwarfs, to the MW and local LIRGs, through to massive high-redshift gas-rich disks.   The disks are initially pressure-supported, but cool rapidly to $<100\,$K and collapse into a wide spectrum of GMCs.   Absent stellar feedback, we find that GMCs undergo a runaway gravitational collapse to high density;  star formation proceeds on approximately a single galaxy averaged dynamical time (Fig. \ref{fig:sfh.ov}), a result that is dramatically inconsistent with observations (Fig.~\ref{fig:ks}).   However, with feedback included, the GMCs dissociate once a modest fraction of their mass has turned into stars and the galaxy develops a turbulent, multi-phase, ISM as long as gas remains.   Quantitatively, the turbulence in the ISM maintains marginal stability to self-gravity, i.e., $Q \sim 1$ (Fig.~\ref{fig:turb.highz.demo}).    Moreover, the galaxies self-regulate and approach a quasi-steady state star formation rate that is consistent with the observed Kennicutt-Schmidt relation over a dynamic range of several orders of magnitude in surface density (Fig.~\ref{fig:ks}).

Our numerical results are reasonably consistent with the observed {global} Kennicutt-Schmidt relation nearly {\em independent} of the high-density star formation law used in the simulation (Fig. \ref{fig:sfh.sflaw} \& \ref{fig:ks}).   This is in contrast to many results in the literature, where free parameters in the high-density star formation law are adjusted to approximately reproduce the Kennicutt-Schmidt relation 
\citep{springel:multiphase,governato04:resolution.fx, dubois:2008.sne.winds.ineff,agertz:disk.fragmentation.model}. 
A weak dependence of the star formation rate on the high-density star formation law is important for developing a predictive galaxy formation model.  It is otherwise difficult to disentangle results that are due to the physics of star formation and/or feedback from those that are due to particular numerical choices/parameters.

 Our star formation model is that  gas turns into stars in dense regions above some threshold density $\rho_{\rm 0}$ at a rate $\dot{\rho}_{\ast}\propto \rho^{n}$.
Varying the normalization of this relation (the high-density star formation efficiency) by a factor of $\sim20$, varying $\rho_{\rm 0}$ by a factor of $\sim 100$, and varying the power-law index $n$ in the 
range $1-2$ changes the quasi-steady state star formation rate by $\lesssim 50 \%$ (Fig. \ref{fig:sfh.sflaw}) (this of course requires that the threshold density be  well-resolved).   Physically, 
this weak dependence arises because the condition for quasi-steady state star formation is that the  momentum injection rate by stellar feedback is sufficient to maintain the ISM at $Q\sim1$.   Reaching $Q \sim 1$ requires a particular turbulent velocity $\delta v$, and thus a particular momentum injection rate, for a given set of global galaxy properties (eq.~\ref{eqn:Q}).    Variations in the high-density star formation law are compensated for by slightly more or less gas collapsing to high densities (and differences in how dense the gas becomes before GMCs are dissociated), so as to produce the same momentum injection rate and hence the same global star formation rate (Fig.~\ref{fig:density.vs.fb}).

The key parameter that determines the efficacy of the stellar feedback in our model is the normalization of the momentum injection rate, $\eta_p$ (eq.~\ref{eqn:mom.vs.L}), where $\dot p \sim (1 + \eta_p \tau_{\rm IR})L/c$;  physically, $\eta_p \lesssim 1$ corresponds to photons leaking out of regions with lower-than-average surface densities while $\eta_p \gtrsim 1$ corresponds to the effects of additional momentum sources (e.g., supernovae and stellar winds) and/or insufficient resolution of the highest optical depth ($\tau_{\rm IR}$) regions.   Numerically, we find less than a factor of 2 change in star formation rate over a factor of $\sim 10$ in $\eta_p$ (Fig.~\ref{fig:sfh.fb}). 
Physically, this is again because maintaining $Q \sim 1$ requires a particular momentum injection rate and thus a particular star formation rate.   Variations in $\eta_p$ are compensated for by the surface densities and thus optical depths $\tau_{\rm IR}$ reached in dense star-forming clumps, maintaining approximately the same momentum injection rate independent of $\eta_p$ (Fig. \ref{fig:sfh.ov} \& \ref{fig:density.vs.fb}).  

Although we have emphasized the importance of momentum input throughout this paper, this is clearly only part of the impact of massive star formation on the ISM of galaxies.  
Which feedback process is the most important depends on the galaxy mass, gas fraction, etc., and on the specific science question of interest.   Heating by photoionization and supernovae, and their effect on molecule formation, are critical physics to include in the formation of the first stars as well in studies of lower-density gas characteristic of dwarf galaxies and the outer parts of more massive disks. 
In the diffuse ISM and the halos of massive galaxies, additional pressure support from cosmic rays and/or magnetic fields may also be important.   
The model presented in this paper is most directly applicable to dense gas in the central kpcs of massive, enriched and evolved systems, in which cooling times are short and molecular fractions are order unity.  
And even in these regions, the model here under-predicts the temperatures of the ``hot'' diffuse ISM 
($T\gtrsim10^{6}\,$K); this gas is likely to be heated by shocks from SNe explosions and fast stellar winds, with $v\sim1000\,{\rm km\,s^{-1}}$.
Although at a given instant, this phase represents only $\sim1\%$ of the gas mass, 
it can have important effects on the generation of galactic super-winds.
In a subsequent paper we will study the combined effect of stellar radiation, stellar winds, and supernovae, with the goal of developing a more widely applicable stellar feedback model for use in galaxy formation.
To extend the study here from idealized disks to disks forming over cosmological timescales, 
it will also be important to incorporate the the realistic cosmological effects of gaseous halos and cold-flow accretion as well as galaxy mergers.

\acknowledgments 
We thank Todd Thompson for helpful discussions and for collaboration that motivated this work. We also 
thank the anonymous referee for a number of thoughtful and useful suggestions. 
Support for PFH was provided by the Miller Institute for Basic Research in Science, University of California Berkeley. EQ is supported in part by the David and Lucile Packard Foundation.   NM is supported in part by the Canada Research Chair program and by NSERC of Canada.\\

\bibliography{/Users/phopkins/Documents/lars_galaxies/papers/ms.bib}

\begin{appendix}

\section{Additional Numerical Tests}
\label{sec:appendix}

\begin{figure}
    \centering
    \plotone{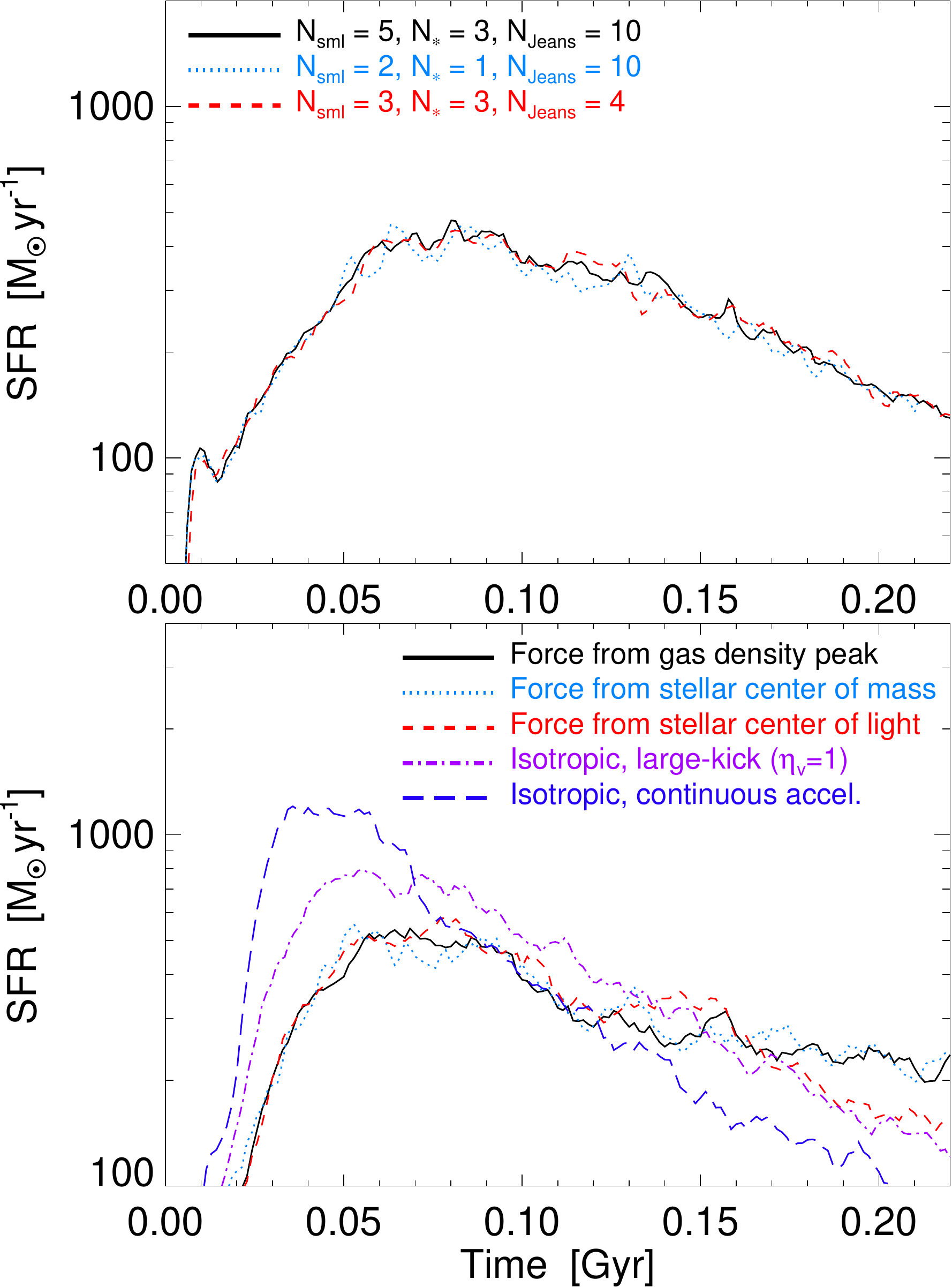}
    \caption{Example star formation histories for the HiZ galaxy model for variations in the details of our numerical method.  {\em Top:} Otherwise identical simulations with 
    different choices of $N_{\rm sml}$ (linking length for clump finder), 
    $N_{\ast}$ (number of lengths to smooth/cut the radiation field), 
    and $N_{\rm Jeans}$ (numerical pressure support to prevent 
    artificial fragmentation). 
    {\em Bottom:} 
    Otherwise identical simulations with variations in the direction in which the feedback is applied. Applying the force radially outwards from the nearest gas density peak, stellar center of mass, or center of light give nearly identical results. 
    The ``isotropic'' models choose the direction of each kick 
    randomly. When individual kicks are relatively rare and large 
    ($\eta_{v}\gtrsim1$), this is somewhat less efficient than radial 
    kicks but still stirs turbulence and slows down star formation.  
    By contrast, in the limit of continuous acceleration, random isotropic forces cancel, 
    impart little net momentum, and the star formation rate is only a factor of $\sim 2$ smaller than that found in the absence of feedback (Fig.~\ref{fig:sfh.ov}).
    \label{fig:sfh.fb.num}}
\end{figure}

In this appendix, we discuss a number of additional numerical tests performed to ensure that 
our conclusions are not sensitive to the precise implementation of the feedback algorithm. 

Figure~\ref{fig:sfh.fb.num} shows examples in which we modified the friends-of-friends search used to identify the nearby density peak that is the origin for our feedback (see \S \ref{sec:stellar.model:clumpids}).  Specifically, we varied 
the parameter $N_{\rm sml}$ between $1-5$  (our standard model adopts $N_{\rm sml}=3$) -- this defines the number of smoothing lengths in which to search for a more-dense gas particle in each iteration.   Within this range we do not see significant differences in most of the identified density peaks; nor is there a significant effect on the star formation history (Fig.~\ref{fig:sfh.fb.num}).
We have also modified the resolution-dependent pressure floor used to avoid artificial collapse, through the parameter $N_{\rm Jeans}$ which represents the minimum number of smoothing lengths which resolve the Jeans length (eq.~\ref{eqn:Pjeans}).  Because all of our feedback-regulated runs have large feedback-induced random velocities, this pressure floor makes no significant difference in these cases (see Fig~\ref{fig:sfh.fb.num}).   It does, however, determine the smallest scale of resolved structure in simulations without feedback \citep[see e.g.][]{robertson:2008.molecular.sflaw}.

To reduce the code run-time and ensure that feedback is relatively local to massive stars, we 
only apply the feedback to gas particles within $N_{\ast}=3$ smoothing lengths of any star particle (\S \ref{sec:stellar.model:momentum}).  We have experimented with this value in the range $2-20$ (the latter value 
including almost all of the gas). Because the optical depths and momentum deposition are 
dominated by the gas closest to the stars, this choice has very small effects on the ISM properties and star formation history (Fig.~\ref{fig:sfh.fb.num}).

One aspect of our method that can have a more significant influence on the results is the direction in which particles are accelerated when the feedback is applied. In the standard model, we accelerate particles away from the gas density peak identified in the friends-of-friends search (\S \ref{sec:stellar.model:momentum}). We have experimented with changing the origin for this force to be the center of mass of the stars or gas or the center of luminosity of the star-forming clump.  
We have also considered models in which the force vector is oriented along the local density gradient (appropriate 
in principle for arbitrary geometries, in the regime where $\tau_{\rm IR}$ falls below unity along that gradient). We find that these models give nearly indistinguishable results, as is shown in Figure~\ref{fig:sfh.fb.num}.   This is essentially because most of the stars are concentrated near the clump center (by any of these metrics) for most of the time when feedback is important.   On small scales in galaxies (or {\em interior} to GMCs), however, the dynamical time can be much shorter than the lifetime of massive stars, and so it is possible that large separations could 
arise between massive stars and the gas from which they formed.   In this case it would be better to 
determine the direction of the force using the local peak in stellar luminosity.

We have also considered experiments in which  we completely ignore the clump density information and kick particles with isotropic, random directions (rather than away from clump centers). 
In our standard model ($\eta_{v}=1$), kicks are somewhat rare but have large 
initial velocities, so the coherent momentum imparted with each kick is still large (even if it is randomly directed).  In this case, the feedback is somewhat less effective than in our standard model and so the star formation rate is somewhat larger  (Fig. ~\ref{fig:sfh.fb.num}).   If the individual kicks particles receive are much smaller (but more frequent), the coherent momentum imparted will be reduced if each is independently randomly oriented. 
As a result, in the limit in which we continuously accelerate particles in random directions (rather than imparting discrete kicks), we find that the feedback has little effect.   The star formation history is similar to models with no feedback (Fig.~\ref{fig:sfh.fb.num}).   This highlights the importance of properly applying the feedback radially away from the center of mass/luminosity of massive star clusters.

There are other aspects of our simulations that are uncertain, independent 
of the stellar feedback model. For example, our cooling function at low temperatures is not exact, 
since we do not explicitly follow chemical networks. We have therefore 
considered various arbitrary changes to the cooling function: setting $\Lambda(T)$ below 
$10^{4}\,$K to a constant median value, or simply forcing all gas at high densities to 
a minimum  temperature $\sim 100$ K.
These introduce $<20\%$ changes in e.g.\, the star formation history, since in all cases the cooling 
time is short relative to the dynamical time. 
On the other hand, removing fine-structure cooling entirely (effectively 
producing a cooling floor at $10^{4}\,$K) dramatically changes the behavior in 
the MW, Sbc, and SMC cases, since this temperature floor is sufficient to artificially prevent collapse to high densities. However it makes little difference in the HiZ case because the 
requirement for $Q>1$ is $c_{s}\gtrsim 30-50\,{\rm km\,s^{-1}}$.

\section{The Effects of Photon ``Leakage''}
\label{sec:appendix:leakage}

\begin{figure}
    \centering
    \plotone{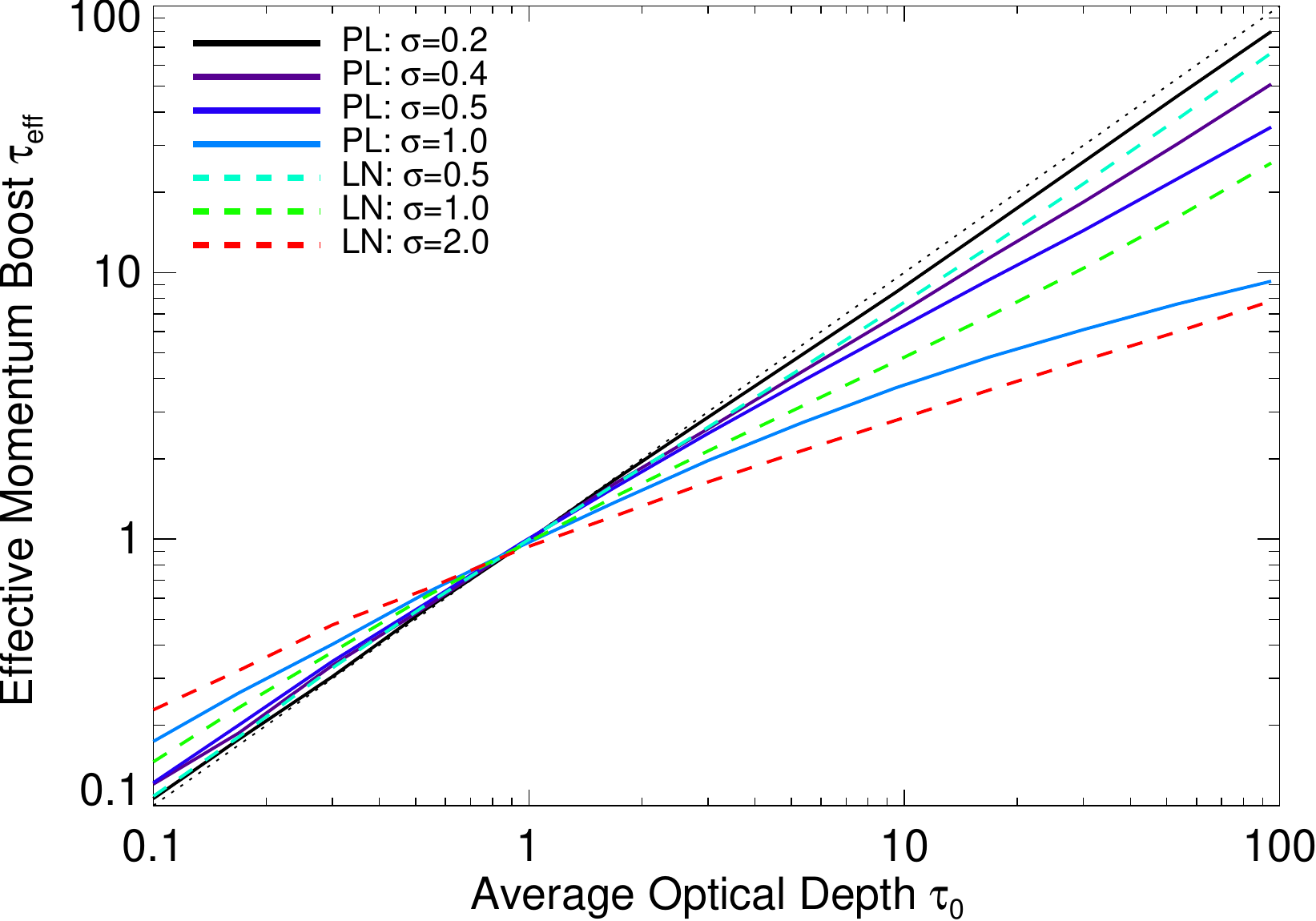}
    \caption{Modified coupling of photon momentum in a self-similar, ``leaky'' medium with a broad 
    density distribution. The true ``boost'' to the coupling $\tau_{\rm eff}$ defined such that 
    $\dot{P} = \tau_{\rm eff}\,L/c$, is plotted as a function of the mean $\tauavg$, for a 
    source at the center of a medium with a random distribution of densities 
    that obeys a power-law (PL; Eq.~\ref{eqn:dPtau.powerlaw}) or lognormal 
    (LN; Eq.~\ref{eqn:dPtau.lognormal}) PDF with logarithmic dispersion $\sigma$. 
    Dotted line shows $\tau_{\rm eff}=\tauavg$, the expectation for a completely homogenous 
    medium. For $\sigma<1$ in the lognormal model, or $<0.5$ in the power-law model, 
    $\tau_{\rm eff}\propto \tauavg$ (with relatively small normalization corrections comparable to 
    small variations in our $\eta_{p}$ parameter). At larger dispersion, 
    the scaling becomes sub-linear, with 
    $\tau_{\rm eff}\propto \tauavg^{1/2\sigma}$ (PL) or 
    $\tau_{\rm eff}\propto \tauavg^{\ln{\tauavg}/4\sigma^{2}}$ (LN). 
    The dispersion in ultra high-res simulations 
    (and observations) corresponds to $\sigma\approx0.5$.  
    \label{fig:taueff.vs.tauavg}}
\end{figure}

\begin{figure}
    \centering
    \plotone{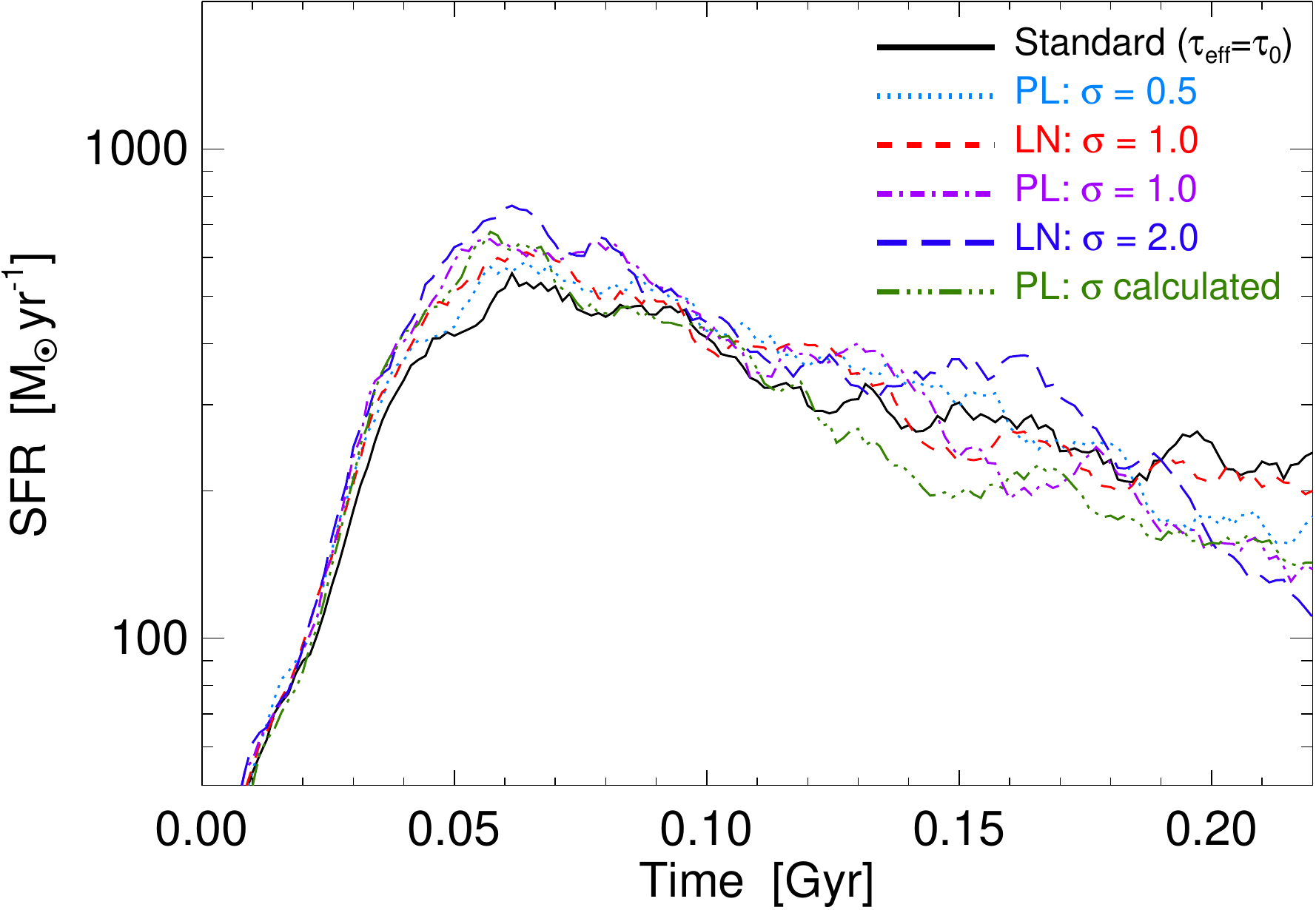}
    \caption{Effects of IR photon ``leakage'' as calculated in Figure~\ref{fig:taueff.vs.tauavg} 
    on the SFH of simulated galaxies (here the HiZ model). We compare our standard 
    model ($\tau_{\rm eff}=\tauavg$) to models using the calculated $\tau_{\rm eff}(\tauavg)$ 
    from Fig.~\ref{fig:taueff.vs.tauavg}, for an assumed universal $\sigma$ and 
    functional form of the density distribution as in that Figure.
    We also compare a model with the power-law/exponential distribution 
    (Eqn.~\ref{eqn:dPtau.powerlaw}) and $\sigma$ calculated on-the-fly 
    as the dispersion in $\ln(\rho)$ within the identified clump radii $R_{\rm clump}$. 
    Larger dispersion in columns leads to more leakage and higher SFRs, but the effect is 
    weak and, in the regime of interest here, similar to a choice of $\eta_{p}$ slightly 
    somewhat less than unity. 
    \label{fig:sfh.vs.leakage}}
\end{figure}

A subtle complication in modeling the effects of radiation pressure 
arises if the ISM is truly inhomogeneous on all scales, including those well below what we model. 
A patch of ISM that appears smooth in the simulations, with some average optical depth $\tauavg$, 
may more accurately (at infinite resolution) exhibit a distribution of local columns, including 
some optically thin lines of sight that could, in principle, allow photons to ``leak out'' at a 
rate much higher than the nominal $\exp(-\tauavg)$ expectation. This would potentially
lower the effective ``boost'' to the radiation pressure from 
$\tauavg\,L/c$ to some $\tau_{\rm eff}\,L/c$. 
It is straightforward to show that leakage will not significantly change the total energy absorbed 
and therefore the IR luminosity density -- once $\tauavg$ is large, it is generically true for 
essentially any realistic distribution of optical depths that {\em most} of the incident optical/UV 
radiation is initially absorbed (whether the escape fraction is a few tens of percent or vanishingly 
small makes only a tens of percent change to $L_{\rm IR}$). 
The concern is rather that IR photons will tend to escape along optically thin sightlines 
before they scatter sufficiently numerous times to impart the full $\sim \tauavg$ momentum 
boost. \citet{krumholz:2009.rad.dom.region.dynamics} 
for example argue that the latter effect means that the 
effective $\tau_{\rm eff}$ can never be larger than a few, even when $\tauavg\gg1$. 
But they do so by assuming an order-unity fraction of un-obscured sightlines, 
independent of $\tauavg$. 
We therefore consider this effect in more detail in this Appendix.

As discussed in the text, the case of a perfectly homogenous density distribution with a 
source at the center is trivial. The opacity along all sightlines is $\tauavg$, so the 
photon scatters an average of $N= \tauavg^{2}$ times as it performs a random walk 
to diffuse out of the sphere. For a random walk, the net momentum flux directed radially 
away from the central source is just $\tau_{\rm eff}\,L/c$ where 
$\tau_{\rm eff} = \sqrt{N}$, which in this case gives $\tau_{\rm eff}= \tauavg$. 

But now consider a case with an inhomogenous density distribution. 
Unfortunately in general, calculating the precise $\tau_{\rm eff}$ for any inhomogenous 
density distribution is complex and cannot be solved analytically -- it requires a full 
radiative transfer solution for each specific density distribution. 
However, we can make considerable progress and obtain reasonably general scaling 
laws with some simplifying assumptions. 
Consider a ``cloud'' of ISM with a well-defined mean $\tauavg $ (which we can 
measure easily in the simulations) enclosing a source at its center; 
for convenience (with no loss of generality) define the cloud radius to be $\ell_{0}=1$.
Now define the ``true'' distribution of optical depths across all sightlines within 
the cloud to be $dP(\tau\,|\, \tauavg)$. 
Finally, assume that the cloud is self-similar with structure on all scales. 
In this limit, the distribution of local densities or 
${\rm d}\tau/{\rm d}\ell$ (where $\ell$ is the distance a long a line of sight) 
is just $dP({\rm d}\tau/{\rm d}\ell) = dP(\tau/\ell_{0}\,|\,\tauavg/\ell_{0})
=dP(\tau\,|\,\tauavg)$. 
We can perform the following relatively simple calculation: 
take a population of photons starting at the center, with initial random directions. 
For each, draw a random ${\rm d}\tau/{\rm d}\ell$ (equivalently, line-of-sight-averaged density), 
and determine in standard monte carlo fashion 
the distance the photon travels before being absorbed 
(for a uniform random variable $p$ between zero and unity, $\Delta \ell = -\ln{(1-p)}/[{\rm d}\tau/{\rm d}\ell]$). 
At each ``scattering,'' determine a new random direction for the photon to be 
re-emitted, and record the locally coupled momentum as the negative of the change in the 
photon momentum. This is iterated until all photons escape the sphere; 
using a large monte-carlo sample of ``photons,'' then, we can determine quantities 
such as the average number of scatterings $\langle N \rangle$ and the 
average momentum imparted (or effective boost $\tau_{\rm eff}$).\footnote{Specifically 
we are interested in the net momentum imparted radially away from the center. As expected, 
all other components of the coupled momentum average to zero.}

Of course, we still need to specify some distribution $dP(\tau\,|\, \tauavg)$. 
Fortunately, we can make a reasonable estimate: 
in ultra-high resolution simulations, we can calculate for example the form of  
$dP(\tau\,|\, \tauavg)$ for each molecular cloud in the simulation, using 
a large number ($\sim1000$) lines-of-sight and integrating the simulation column 
along each sightline. In \S~\ref{sec:stellar.model:momentum}, we discuss this process and note that the 
resulting column density distribution on a per-cloud basis can be well-approximated by a 
near-universal function 
\be
\label{eqn:dPtau.powerlaw}
dP(\tau\,|\, \tauavg) \approx \frac{1}{2\,\sigma}\,
\exp{{\Bigl(}-\frac{|\ln{(\tau/\tauavg)}|}{\sigma}{\Bigr)}}\,
\frac{d\tau}{\tau}
\ee
with $\sigma$ (the standard deviation) ranging from $0.25-1.0$ 
($0.1-0.4$\,dex) with a median $\sigma=0.5$ ($0.22$\,dex). 
This is very similar to the distribution of columns estimated in 
much higher resolution simulations of individual GMCs and sub-cloud clumps 
\citep[often with very different physics included; see e.g.][]{ostriker:2001.gmc.column.dist}, and 
to observational estimates of the column density distribution across observed 
GMCs \citep{wong:2008.gmc.column.dist,goodman:2009.gmc.column.dist}. 

Note that this distribution is exponential in $\log(\tau)$ -- really, the key 
behavior is that the number sightlines at small or large 
$\tau$ falls off as a {\em power-law} in $(\tau/\tauavg)$, rather than 
an exponential or log-normal. This is important, as we will see below.

Given some $\sigma$, then, it is straightforward to perform the 
Monte Carlo calculation of $\tau_{\rm eff}$ described above.
Figure~\ref{fig:taueff.vs.tauavg} shows the coupled momentum 
$\tau_{\rm eff}$ as a function of the average $\tau_{0}$, for a few values of 
$\sigma$. 
At very low $\tauavg$, the coupled boost drops off rapidly because a large 
fraction of sightlines are optically thin -- but in this case, the ``boost'' is negligible 
in either case (with or without leakage). 
At low to moderate $\tauavg$, the effective $\tau_{\rm eff}$ rises with $\tau$ in a linear 
fashion as we would naively expect. 
At high-$\tauavg$, however, the behavior depends on $\sigma$, 
with a critical change in behavior around $\sigma=1/2$. This can be understood from 
the form of $dP(\tau\,|\, \tauavg)$.

For the distribution in Equation~\ref{eqn:dPtau.powerlaw}, 
the fraction of optically-thin sightlines ($\tau < 1$) 
scales simply as 
\be
f_{\rm thin} = \frac{1}{2\,\sigma}\,\tauavg^{-1/\sigma}
\ee
So the average number of scatterings needed before a photon will ``find'' an optically 
thin sightline -- and therefore 
have a high probability of escaping the cloud -- 
is, crudely, $N_{\rm thin}\sim 1/f_{\rm thin} \propto \tauavg^{1/\sigma}$. 
On the other hand, the number of scatterings needed before 
the photon would diffuse out of the cloud assuming it did {\em not} find 
an optically thin sightline is just 
$N_{\rm diffuse}\sim \tauavg^{2}$. 
What matters for the behavior at high-$\tauavg$ is just this power-law falloff (exactly 
how we model the ``core'' and high-$\tau$ part of the PDF make almost no difference 
to the values of $\tau_{\rm eff}$ plotted in Figure~\ref{fig:taueff.vs.tauavg} at high-$\tauavg$). 

If $\sigma<1/2$, then, $N_{\rm thin}$ grows more rapidly than $N_{\rm diffuse}$; 
in other words, the number of optically-thin sightlines falls off sufficiently rapidly 
at large $\tauavg$ that the typical photon will undergo its expected number of 
$\sim \tauavg^{2}$ scatterings to diffuse out before it ``leaks'' out of an optically 
thin sightline, so that the coupled momentum $\tau_{\rm eff} \propto \sqrt{N} \propto \tauavg$. 
There is a linear normalization correction to $\tau_{\rm eff}$, which we can 
estimate analytically by considering the average distance travelled between 
scatterings, i.e.\ $\langle \Delta \ell \rangle \propto \langle \tau^{-1} \rangle$ -- 
this accounts for the fact that, on average, slightly more thin or thick sightlines allow for 
fewer or more scatterings before escape: 
we obtain the result that, for $\tauavg\gg1$, $\tau_{\rm eff} \rightarrow (1-\sigma^{2})\,\tauavg$. 

If $\sigma>1/2$, however, then $N_{\rm diffuse}$ grows more rapidly than $N_{\rm thin}$; 
so the average photon will ``leak out'' after just $N_{\rm thin}$ scatterings, before 
it can couple the full $\tauavg \sim \sqrt{N_{\rm diffuse}}$ ``boost.'' 
The actual coupled momentum should instead scale as $\tau_{\rm eff} \sim \sqrt{N_{\rm thin}}$, 
which gives $\tau_{\rm eff} \propto \tauavg^{1/2\sigma}$. 
Again, we can analytically estimate the pre-factor for $\tauavg\gg1$, and obtain 
$\tau_{\rm eff}\rightarrow \sqrt{\sigma/2\,\Gamma[1/\sigma]}\, \tauavg^{1/2\sigma}$. 
The important point is that in this regime, the scaling is {\em sub-linear} in $\tauavg$. 
There is still an approximately linear regime at moderate $\tauavg$, but 
for very high $\tauavg$, the ``boost'' becomes more limited. 

We stress that this behavior arises because the assumed $dP(\tau\,|\,\tauavg)$ behaves as a 
{\em power-law} at low $\tau/\tauavg$ -- in other words, this allows for essentially the 
maximal ``long tail'' of low-$\tau$ sightlines towards a central source with high average optical 
depth. 
Since $dP({\rm d}\tau/{\rm d}\ell\,|\,\tauavg/\ell_{0})$ reflects the local three-dimensional 
density distribution, it might for example be more natural to assume it should have a 
lognormal form: 
\be
\label{eqn:dPtau.lognormal}
dP(\tau\,|\, \tauavg) \approx \frac{1}{\sigma\,\sqrt{2\pi}}\,
\exp{{\Bigl(}-\frac{\ln{(\tau/\tauavg)}^{2}}{2\,\sigma^{2}}{\Bigr)}}\,
\frac{d\tau}{\tau}
\ee
The results of assuming this distribution are shown in Figure~\ref{fig:sfh.vs.leakage}. 
Given this distribution, the high-$\tauavg$ limit essentially {always} 
gives a linear scaling $\tau_{\rm eff} \propto \tauavg$. 
The reason is obvious given the arguments above -- for a lognormal, 
the fraction of low-$\tau$ sightlines will decline much faster than a power-law, 
so independent of $\sigma$, the probability of finding optically thin sightlines 
at $\tau\ll \tauavg$ will fall much faster than $\tauavg^{-2}$. 
For $\sigma\lesssim1$, then, $\tau_{\rm eff}\rightarrow \tauavg$ when $\tauavg\gg1$. 
At very large $\sigma$ or more moderate $\tauavg$, of course, $\tau\sim1$ may still fall within 
the ``core'' of the lognormal. For Equation~\ref{eqn:dPtau.lognormal}, 
the fraction of optically thin sightlines when $\tauavg\gg1$ is 
$f_{\rm thin} = (\sigma/\sqrt{2\pi})\,\ln^{-1}{(\tauavg)}\,\tauavg^{-(\ln{\tauavg}) / (2\sigma^{2}) }$, 
or $f_{\rm thin}\propto \tauavg^{- (\ln\tauavg) / (2\sigma^{2}) }$. 
The requirement that this fraction drop faster than $\tauavg^{-2}$ 
(to give $\tau_{\rm eff}\sim\tauavg$) therfore 
becomes $\tauavg \gtrsim \exp{(4\,\sigma^{2})}$. 
For the reasonable values of $\sigma$ up to order unity, this is easily satisfied for 
high-$\tauavg$. But if $\sigma$ were very large (say $\sim2$), this rapidly becomes 
extremely large, and so we return to the 
$N_{\rm thin} < N_{\rm diffuse}$ limit, and 
obtain the sub-linear scaling 
$\tau_{\rm eff} \rightarrow (2\pi)^{1/4}\,[\ln{(\tauavg)}/\sigma]^{1/2}\,\tauavg^{(\ln{\tauavg})/4\sigma^{2}}$. 
Even in this regime, however, it is worth noting that at very high 
$\tauavg$, the power-law model of Equation~\ref{eqn:dPtau.powerlaw} still has a 
larger optically thin fraction -- for a fixed $\sigma$, that model gives a maximal 
effect of leakage.

We can test the effects of this in our simulations by replacing 
the standard boost of $\tauavg$ with one of the appropriate $\tau_{\rm eff}$ 
calculated above with some fixed $\sigma$ 
(using the curves in Fig~\ref{fig:taueff.vs.tauavg} to define an interpolation table). 
Obviously, for either distribution, a value of $\sigma<0.5$ will make no difference to 
any of our conclusions because $\tau_{\rm eff}\propto \tauavg$: the normalization correction is 
completely equivalent to variations in $\eta_{p}$, discussed in the text 
(and in the regime of very low $\tau$, the boost scaling is not important). 
And for a lognormal distribution, any $\sigma<1$ will yield identical results. 
We therefore consider experiments with 
the exponential/power-law distribution and assumed $\sigma=0.5,\,1.0$ and 
lognormal distribution with assumed $\sigma=1.0,\,2.0$. 
For the power-law distribution with $\sigma=0.5$ or lognormal with $\sigma=1.0$, 
$\tau_{\rm eff}$ begins to deviate from $\tauavg$ at $\tauavg\gg1$, but the differences 
are sufficiently small that we do not see a large effect (they are roughly comparable 
to choices of $\eta_{p}=2/3$ and $=1/2$, respectively, 
and so change the expected SFRs only at the $20-30\%$ level). 
For the very large choices of $\sigma$, however, we do expect and see some deviations. 
The equilibrium SFR is systematically larger by a factor of $\sim2$, similar to a 
small $\eta_{p}\sim1/4$ (since the median $\tau_{\rm IR}\sim30-50$ becomes 
$\tau_{\rm eff}\sim10$ here, this is expected). 
Visual inspection in these cases also confirms there are some small sub-regions 
in the galaxy nucleus where the gas consumption is near-runaway (these do not contain 
much of the mass, but they have the highest densities, $\tau_{\rm IR}\sim100$). 
Finally, we have also considered runs in which the power-law model is adopted, 
but with $\sigma$ taken from the local gas properties. Specifically, we take all of the gas inside the 
identified clump radius $R_{\rm clump}$, and (knowing the density of each particle) 
compute the dispersion in $\ln{(\rho)}$ which we use as $\sigma$. We also add in quadrature 
a minimum $\sigma=0.25$ ($0.1\,$dex) which is about the minimum dispersion we see 
in ultra-high resolution simulations (in order to again be conservative and allow for 
significant leakage). The results of this run are quite similar to our 
default $\tau_{\rm eff}=\tauavg$ model and/or the $\sigma=0.5$ model, which we expect 
since, as noted in the text, the typical $\sigma$ we measure in simulation clumps is about 
$0.5$. It is worth noting though, that the typical $\sigma$ increases as we consider older and 
older stars, as a consequence of feedback in earlier stages driving out gas and ``punching holes'' 
in the gas distribution. Of course, the mean optical depth is also going down here, 
and we saw in the text that stars with the youngest ages $\lesssim1\,$Myr dominate the 
radiative momentum input. So accounting for leakage accelerates the rate at which old 
stars luminosity can escape without coupling in the IR, but does not change our conclusions. 

We should also emphasize that other sub-grid effects could in fact {\em raise} 
$\tau_{\rm eff}$ at fixed $\tauavg$. This includes some non-trivial geometric cases 
where photons can be more efficiently trapped. Also, recall that we define 
$\tauavg$ in the model as the globally averaged $\tau$ out to a given radius 
$\propto M_{\rm enc}\,R^{-2}$, rather than the line-of-sight integrated $\tau$. 
If the gas within $R$ is distributed with any average density profile that rises towards the 
stars, the appropriate $\tau_{\rm eff}$ should be larger than that in 
Fig.~\ref{fig:taueff.vs.tauavg} (for a pure power-law profile 
 $\rho\propto r^{-\alpha}$, this gives a factor $\propto 1/(1-\alpha)$ which is 
 actually divergent for $\alpha>1$). In fact, if we calculate the true median line-of-sight 
 integrated $\tau$ in our ultra-high resolution simulations and compare it to the adopted 
 $\tauavg$, the typical correction would amount to a ``boost'' of $\eta_{p}\approx2$. 
And if we under-resolve collapse such that the gas ``should'' collapse a factor 
$\sim \psi$ further in radius than our resolution limit allows, then $\tau_{\rm eff}\sim \psi^{2}\,\tauavg$ 
would be appropriate. It is difficult, therefore, to identify a ``more accurate'' 
model than our $\tau_{\rm eff}\propto \tauavg$ that is robust at the factor $\sim2$ level. 

We therefore conclude that photon leakage is unlikely to qualitatively 
change our conclusions, given observationally and theoretically realistic distributions of 
column densities towards optically thick sources. 
However, it might be important in the most dense systems observed: starburst nuclei and AGN. 
The average IR optical depths in these regions can reach values $>100$. 
The absolute value of the correction to $\tau_{\rm eff}$ here could therefore be quite large -- a factor $\sim10$ 
rather than $\sim2$, if $\sigma$ is sufficiently large. Moreover, the sub-linear behavior 
of $\tau_{\rm eff}$ could be very important, because these regions both have high-$\tauavg$ 
and have dynamical times that are short relative to the stellar evolution timescale. 
In this joint limit, the luminosity required to support the system is 
$\tau_{\rm eff}\,L/M \propto (M/R^{2})$ or $L/M \propto \tauavg/\tau_{\rm eff}$. 
When we are in the linear regime ($\tau_{\rm eff}\propto \tauavg$), either because of 
low $\sigma$ or low $\tauavg\lesssim10$, this implies that the system can self-regulate 
on both small and large scales once a fraction (a few percent) of the mass becomes stars. 
However, if $\tau_{\rm eff}$ is significantly sub-linear, than the $L/M$ needed to stabilize is a 
rising function of $\tau$ -- in other words, the system is vulnerable to runaway collapse 
\citep{fall:2010.sf.eff.vs.surfacedensity}.
Such collapse could be quite interesting in these regions, however, since it would proceed with 
regions above a critical $\tauavg$ running away to turn entirely into stars, while neighboring regions 
that had smaller $\tauavg$ do not collapse -- and once the global $L/M$ reached a given threshold, 
the low-density regions would be self-regulated. One might imagine a regime of global 
self-regulation on these scales, but without {\em local} self-regulation, perhaps making these 
regimes particularly interesting for the formation of globular clusters, 
dwarf galaxy nuclei, and/or super star clusters.

\end{appendix}
\end{document}